\documentclass[
 reprint,
 amsmath,amssymb,
 prl
]{revtex4-2}

\usepackage[ngerman,american]{babel} 
\usepackage{graphicx}
\usepackage{siunitx}
\usepackage{diffcoeff} 
\usepackage{xspace} 
\usepackage{csquotes}
\usepackage[hidelinks]{hyperref} 
\usepackage[nosort]{cleveref}

\newcommand*{\sep}{\ensuremath{\text{sep}}\xspace}
\newcommand*{\rest}{\ensuremath{\text{rest}}\xspace}

\newcommand*{\spike}{\ensuremath{\text{sp}}\xspace}
\newcommand*{\tspike}{\ensuremath{t_\spike}\xspace}

\newcommand*{\taus}{\ensuremath{\tau_\text{s}}\xspace}
\newcommand*{\taum}{\ensuremath{\tau_\text{m}}\xspace}

\crefname{section}{Sec.}{Secs.}
\crefname{equation}{Eq.}{Eqs.}
\crefname{figure}{Fig.}{Figs.}

\begin{document}

\title{Smooth Exact Gradient Descent Learning in Spiking Neural Networks}

\author{Christian Klos}
 \email{cklos@uni-bonn.de}
\author{Raoul-Martin Memmesheimer}
 \email{rm.memmesheimer@uni-bonn.de}
\affiliation{Neural Network Dynamics and Computation, Institute of Genetics, University of Bonn, 53115 Bonn, Germany}


\begin{abstract}
Gradient descent prevails in artificial neural network training, but seems inept for spiking neural networks as small parameter changes can cause sudden, disruptive (dis-)appearances of spikes. Here, we demonstrate exact gradient descent based on continuously changing spiking dynamics. These are generated by neuron models whose spikes vanish and appear at the end of a trial, where it cannot influence subsequent dynamics. This also enables gradient-based spike addition and removal. We illustrate our scheme with various tasks and setups, including recurrent and deep, initially silent networks.
\end{abstract}

\maketitle

\paragraph{Introduction.}
Biological neurons communicate via short electrical impulses, called spikes~\cite{DA01}. Besides their overall rate of occurrence, the precise timing of single spikes often carries salient information~\cite{gollisch2008rapid,wolfe2010sparse,saal2016importance,sober2018millisecond}. Taking into account spikes is therefore essential for the modeling and the subsequent understanding of biological neural networks~\cite{DA01,gerstner2014neuronal}. To build appropriate spiking network models, powerful and well interpretable learning algorithms are needed. They are further required for neuromorphic computing, an aspiring field that develops spiking artificial neural hardware to apply them in machine learning. It aims to exploit properties of spikes such as event-based, parallel operation (neurons only need to be updated when they send or receive spikes) and the temporal and spatial (i.e.~in terms of interacting neurons) sparsity of communication to achieve tasks with unprecedented energy efficiency and speed~\cite{Pfeiffer2018,roy2019spikebased,schuman2022opportunities}.

The prevalent approach for learning in non-spiking neural network models is to perform gradient descent on a loss function~\cite{goodfellow2016deep,kriegeskorte2019neural}. 
Importantly, during such learning the representations change continuously and in a predictable manner as the networks are compositions of functions that are continuous in the network parameters.
The transfer of gradient descent learning to spiking networks is, however, problematic due to the all-or-none character of spikes: 
The (dis-)appearance of spikes is not predictable from gradients computed for nearby parameter values. This is because the gradient only accounts for changes in spike timing, of those spikes present when it is computed.
Thus, a systematic addition or removal of spikes via exact gradient descent is seemingly not possible. This can, for example, lead to permanently silent, so-called dead neurons~\cite{eshraghian2023training,Taherkhani2020review} and to diverging gradients~\cite{booij2005gradient}. Further, the network dynamics after a spike (dis-)appearance may change in a disruptive manner~\cite{Vreeswijk1998,JMT08,MW12,memmesheimer2014learning}. This can result in discontinuous changes of the representations, which are given by the spike times, and of the loss during learning.

Nevertheless, there are two popular approaches for learning in spiking neural networks based on gradient descent: The first approach, surrogate gradient descent, assumes binned time and replaces the binary activation function with a continuous-valued surrogate for the computation of the gradient~\cite{Neftci2019surrogate}. It thus sacrifices the crucial advantage of event-based processing and learning only from spikes and necessitates the computation of state variables in each time step as well as their storage~\cite{wunderlich2021eventbased} (but see \cite{perez-nieves2021sparse}). Furthermore, the computed surrogate gradient is only an approximation of the true gradient. The second approach, spike-based gradient descent, computes the exact gradient of the loss by considering the times of existing spikes as functions of the learnable parameters~\cite{BKP02,eshraghian2023training}. It allows for event-based processing but relies on ad-hoc measures to deal with spike (dis-)appearances and gradient divergence, in particular to avoid dead neurons~\cite{Comsa2020snnalpha,Göltz2021neuromorphic,Mostafa2018temporal,nowotny2022loss,Kheradpisheh2020s4nn}.

Here we show that disruptive (dis-)appearances of spikes can be avoided. Consequently, all network spike times vary continuously and in some network models even smoothly, i.e.~continuously differentiably, with the network parameters. This allows us to perform non-disruptive, exact gradient descent learning, including, as we show, the systematic addition or removal of spikes.

\begin{figure}
\centering
\includegraphics{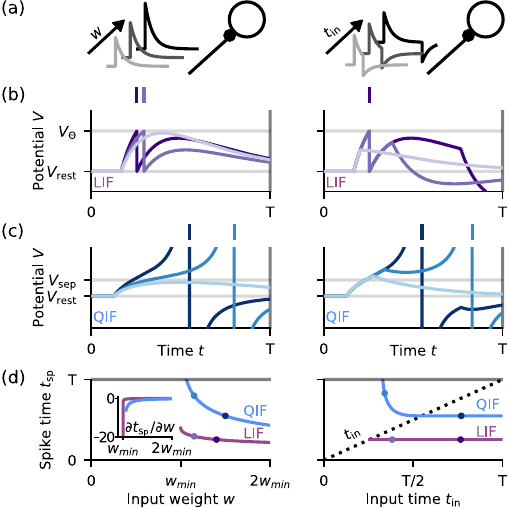}
\caption{\label{fig:fig1} Disruptive and non-disruptive appearance of spikes. 
(a,b,d) Spike times of the LIF can appear disruptively, in the middle of a trial. (a,c,d) Spike times of the QIF only appear non-disruptively at the trial end and otherwise change continuously with changed parameters.
Left column: a neuron receives a single input, whose weight is increased (traces with increasing saturation). Right column: a neuron receives an excitatory as well as an inhibitory input whose arrival is moved to larger times.
(a) Setup (gray: different input currents). (b) LIF membrane potentials (purple traces, saturation corresponding to (a); $V_\rest,V_\Theta$: resting and threshold potential; $T$: trial duration) and spikes (top, ticks). (c) Like (b) for the QIF ($V_\sep$: separatrix potential). (d) Times of the first output spike as function of the changed parameter; dots correspond to equally colored spikes in (b,c). Left: $w_\text{min}$: weight at which the spike appears, at finite time for the LIF, at infinity for the QIF. Inset: Spike time gradient, divergent for LIF.}
\end{figure}

\paragraph{Neuron model.}
The most frequently employed neuron models when learning spiking networks are variants of the leaky integrate-and-fire neuron (LIF)\cite{Supp,Zenke2018,memmesheimer2014learning,Neftci2019surrogate,wunderlich2021eventbased,Göltz2021neuromorphic,cramer2022surrogateneuromorph,perez-nieves2021sparse,Comsa2020snnalpha,nowotny2022loss,BKP02,booij2005gradient}. LIFs, however, suffer from the aforementioned disruptive spike (dis-)appearance. For example, spikes can appear in the middle of a trial due to a continuous, arbitrarily small change of an input weight or time (\cref{fig:fig1}a,b). Here and in the following, a trial refers to an individual run of an experiment with finite duration.

We therefore consider instead another important standard spiking neuron model, a quadratic integrate-and-fire neuron (QIF)~\cite{latham2000intrinsic,I07,gerstner2014neuronal}. Its membrane potential dynamics are governed by $\dot{V}=V(V-1)+I$, where $I$ consists of temporally extended, exponentially decaying synaptic input currents, $\taus\dot{I}=-I+\taus\sum_iw_i\sum_{t_i}\delta(t-t_i)$. Here, $\taus$ is the synaptic time constant, measured in multiples of the membrane time constant $\taum=1$, and $i$ indexes the presynaptic neurons, which spike at times $t_i$ and have a synaptic weight $w_i$. In contrast to the LIF, where $\dot{V}$ decays linearly with $V$, the QIF explicitly incorporates the fact that in biological neurons the membrane potential further increases due to a self-amplification mechanism once it is large enough. As this generates spike upstrokes, the QIF may be considered as the simplest truly spiking neuron model~\cite{I07}. The voltage self-amplification is so strong that the voltage actually reaches infinity in finite time. One can define the time when this happens as the time of the spike, reset and onset of synaptic transmission. We adopt this and henceforth call positive infinity the threshold of the QIF for simplicity. For sufficiently negative voltage, the voltage increases strongly as well. The neuron can thus be reset to negative infinity, from where it quickly recovers. For LIFs one needs to define finite threshold and reset potentials.

\paragraph{Non-disruptive (dis-)appearance of spikes and smooth spike timing.}
In the QIF, spike times only (dis-)appear at the trial end; otherwise they change smoothly with the network parameters. Importantly, this kind of spike (dis-)appearance is non-disruptive, since the are no more spiking dynamics after the trial end that could be affected.

The mechanism underlying this feature can be intuitively understood: The voltage slope $\dot{V}$ at the threshold is infinitely large. If there is a small change for example in an input weight (\cref{fig:fig1} left column, blue curves), $V$ and $\dot{V}$ will still be large close to where the spike has previously been. Therefore a spike will still be generated, only a bit earlier or later, unless it crosses the trial end. This is in contrast to the LIF, where $\dot{V}$ at the threshold can tend to zero and a spike can therefore abruptly (dis-)appear, accompanied by a diverging gradient (\cref{fig:fig1} left column, purple curves). A similar mechanism applies if there are changes in an input time as in \cref{fig:fig1} right column: An inhibitory input is moved backward in time until it crosses the time of an output spike generated by a sole, previous excitatory input ($t_\text{in}$ crosses \tspike in \cref{fig:fig1}d right). In the QIF $V$ and $\dot{V}$ are infinitely large at this point, such that the additional inhibitory input is negligible compared to the intrinsic drive. Thus there is no abrupt change in spike timing. In contrast, in the LIF the inhibitory input induces a downward slope in the potential also if it is at the threshold. The spike induced by the excitatory input alone therefore suddenly appears once the inhibitory input arrives later.

In Supplemental Material Sec.~II~\cite{Supp}, we prove the smoothness of the spike times and their non-disruptive (dis-)appearance in the general case with multiple inputs and output spikes.

\paragraph{Pseudodynamics and pseudospikes.}
The non-disruptive disappearance of spikes allows spike-based gradient descent to remove them in a controlled manner, by shifting them past the trial end. In contrast, the gradient contains no information about spike appearances at the trial end, precluding the systematic addition of spikes. Being able to add spikes is, however, important because a neuron may initially or at some point during learning spike insufficiently often for the task or even be completely silent.

To solve this problem, we appropriately extend the ordinary dynamics by what we call pseudodynamics.
Concretely, we propose two types of pseudodynamics. 
In the first type, which we use in our applications, the neurons continue to evolve as QIFs, but with an added constant, suprathreshold drive, until they have spiked sufficiently often for the task~\cite{Supp}. We call the additional spikes pseudospikes. They only affect the pseudodynamics of postsynaptic neurons, by controlling the value of the added drive. This ensures generically non-zero gradients.
The continued evolution as a QIF ensures continuity and mostly smoothness of the spike times, even if a spike transitions from a pseudospike to an ordinary one.
In Supplementary Material Sec.~IB~\cite{Supp} we suggest a second approach where the spike times remain completely smooth.

Both types of pseudospike times have several useful properties: (i) They depend continuously and mostly smoothly on the network parameters, also when the pseudospikes cross the trial end to turn into ordinary spikes. (ii) If the voltage at the trial end increases, the pseudospike times decrease, intuitively because the neuron is already closer to spike. (iii) Pseudospikes affect postsynaptic pseudospikes but not ordinary ones. (iv) The pseudospikes interact such that the components of the gradient in multi-layer networks are generically non-zero also if neurons are inactive during the actual trial duration. (v) The pseudospike times are computable in closed form.

Similar pseudospike time functions can be found for other neuron and synapse models with continuous spike times such as QIFs with infinitesimally short synaptic currents that generate voltage jumps~\cite{Supp}.

\paragraph{\label{sec:apps}Gradient descent learning.}

In the following, we apply spike-based gradient descent learning on the neural network models with continuous spike times identified above. We choose single neuron models with a closed-form solution between spikes and for the time of an upcoming spike. The former enables and the latter simplifies the use of efficient event-based simulations and modern automatic differentiation libraries~\cite{bradbury2018jax,engelken2023sparseprop}.

Interestingly, such solutions exist for the QIF with temporally extended, exponentially decaying synaptic input currents if $\taus=\taum/2$~\cite{Supp}. This is compatible with often assumed biologically plausible values, for example with a membrane time constant about \qty{10}{\ms} and a synaptic time constant about \qty{5}{\ms}~\cite{gerstner2014neuronal,DA01}. In the examples in this article, we therefore use these values. 

In the last of our three applications we employ oscillating QIFs with infinitesimally short input currents. Between spikes, they evolve with a constant rate of change in an appropriate phase representation~\cite{EK86,latham2000intrinsic,I07,gerstner2014neuronal}, which further simplifies the event-based simulations. While their spike times are continuous, they are not smooth, as the derivative with respect to the time or weight of an input spike time jumps if it crosses another one.

Code to reproduce the results is publicly available~\cite{klos2024code}.

\paragraph{\label{sec:apps_single}Single neuron learning.}

\begin{figure}
\centering
\includegraphics{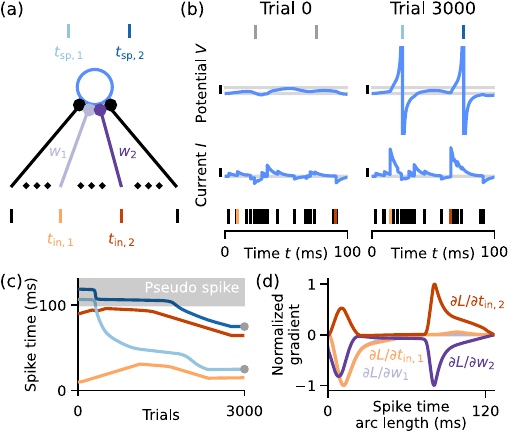}
\caption{\label{fig:fig2} Smooth gradient descent learning in a QIF. 
(a) Weights (purple) and times (orange) of two inputs are learned to adjust the first two output spike times (blue).
(b) Left: Before learning, the neuron does not spike (gray ticks: target spike times; horizontal gray lines: $V_\sep$, $V_\rest$ or zero input current; black bars: potential or current difference of one; orange, black ticks: learned, other input spikes). Right: After learning, the neuron spikes at the desired times (blue ticks cover gray ticks). 
(c) During learning, the (pseudo (gray area)) spike times change smoothly (colors as in (a); gray circles: target spike times). 
(d) The components of the gradient of the loss function $L$ change continuously during learning ($\difsp{L}{w_1}$ mostly covered by $\difsp{L}{t_{\text{in},1}}$). Learning progress is displayed as a function of the arc length of the output spike time trajectories since the start of learning~\cite{Supp}.}
\end{figure}

As a first illustration of our scheme, we learn spike times of a single QIF (\cref{fig:fig2}a, see \cite{Supp} for details on models and tasks). Initially it does not spike at all during the trial (\cref{fig:fig2}b, left). We apply spike-based gradient descent to minimize the quadratic difference between two target and the first two output spike times (which may also be pseudospike times). The neuron is set to initially generate two pseudospikes, one for each target spike time. While not necessary in the displayed task, superfluous (pseudo-)spikes can be included into the loss function with target behind the trial end, to induce their removal if they enter the trial.

The use of pseudospikes allows to activate the initially silent neuron (\cref{fig:fig2}c, gray background). In doing so, the pseudospike times transition smoothly into ordinary spike times (\cref{fig:fig2}c, white background). They are then shifted further until they lie precisely at the desired position on the time axis (\cref{fig:fig2}b, right). The spike times change smoothly (\cref{fig:fig2}c) and the loss gradient is continuous (\cref{fig:fig2}d). The example illustrates that our scheme allows to learn precisely timed spikes of a single neuron -- in a smooth fashion and even if the neuron is initially silent.

\paragraph{\label{sec:apps_rnn}Learning a recurrent neural network.}

\begin{figure}
\centering
\includegraphics{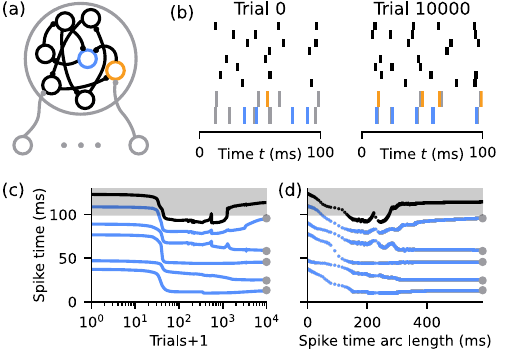}
\caption{\label{fig:fig3} Learning precise spikes in an RNN. (a) Network schematic. Neurons receive in each trial the same spikes from external input neurons (gray). Spike times of the first two network neurons are learned (blue and orange). 
(b) Spikes of network neurons before (left) and after (right) learning (colored ticks: spikes of first two neurons; gray ticks: target times).
(c) Spike time trajectories of the first neuron during learning. Desired spikes (blue traces) shift towards their target times (gray circles). The first superfluous spike (black trace) is pushed out of the trial. (Gray area indicates pseudospikes.)
(d) Same as (c) but the spike times are shown as a function of the arc length of the output spike time trajectories~\cite{Supp}, which demonstrates their continuity, despite the occurrence of large gradients (cf.~the step-like change in (c)).}
\end{figure}

Next, we consider the training of a recurrent neural network (RNN), where spike time changes have a global impact. It can be useful for the reconstruction of cortical networks \cite{memmesheimer2014learning,kim18learning,das2020reconstruction}.
We consider a fully-connected RNN of ten QIFs with 
external inputs and learn the spike times of two network neurons by updating the recurrent weights and initial conditions (\cref{fig:fig3}a).
In contrast to the learning of all network spikes~\cite{MT06A,memmesheimer2014learning}, this does not reduce to independently finding a mapping from given input spike times to output spike times for each neuron.

\begin{figure}
\centering
\includegraphics{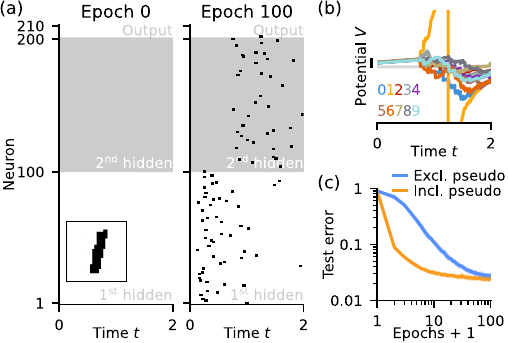}
\caption{\label{fig:fig4} 
MNIST task.
(a) Spike raster plot of the three-layer network. Left: Silent neurons before learning. Inset: example input also used on the right and in (b). Right: Sparse spiking after learning. 
(b) Voltage dynamics of the output neurons after learning (horizontal gray line: $V_\rest$; black bar: potential difference of one).
(c) Classification error dynamics. Utilizing pseudospikes also during testing (orange) generates smaller test errors in early training (solid lines indicate mean and shaded areas std over ten network instances). 
}
\end{figure}

Our scheme is successful also in this scenario and the spike times are precisely learned (\cref{fig:fig3}b). As in the previous example, the spike times of the first neuron change continuously during learning without discrete jumps (\cref{fig:fig3}c,d). Due to large gradients, which are typical for all kinds of RNNs~\cite{pascanu13rnn}, some changes are 
jump-like for the second neuron~\cite{Supp}. 
The underlying 
continuity 
becomes clear when restricting the maximal spike time change per step using
adjustable update step sizes~\cite{Supp}. Hence, this example illustrates the applicability of our scheme to recurrent networks.

\paragraph{\label{sec:apps_mnist}Standard machine learning task.}

Finally, we apply our scheme to the classification of hand-written single-digit numbers from the MNIST dataset, which is a widely used benchmark in neuromorphic computing (e.g.~\cite{wunderlich2021eventbased,Göltz2021neuromorphic,cramer2022surrogateneuromorph}). 

We employ a three-layer feedforward network consisting of oscillatory QIFs with infinitesimally short input currents. For each input pixel, there is a corresponding input neuron, which spikes once at the beginning of the trial if the binarized pixel intensity is one and otherwise remains silent. 
The index of the neuron in the output layer that spikes first is the model prediction~\cite{thorpe2001rapid}. 

To demonstrate that our scheme allows to solve the dead neuron problem even if neurons in multiple layers are silent, we randomly initialize network parameters such that there are initially basically no ordinary spikes (\cref{fig:fig4}a, left).  
Yet, the pseudospike time-dependent, imposed interaction between the neurons allows to backpropagate errors. Hence, the hidden and output neurons are activated (\cref{fig:fig4}a right,b). 
Finally basically all hidden neurons spike before the first output spike for some input image~\cite{Supp}, indicating that they contribute to inference. Still activity is sparse.
The final accuracy of \qty{97.3}{\percent} when only considering ordinary output spikes is comparable to previous results with similar setups~\cite{Mostafa2018temporal,Comsa2020snnalpha,Göltz2021neuromorphic,sakemi2023snntemporal}. If we also allow pseudospikes
during testing (which only affects trials without ordinary output spikes), the accuracy does not change much. The minimal error level is, however, reached faster (\cref{fig:fig4}c). Thus, our scheme achieves competitive performance in a neuromorphic benchmark task even if almost no neuron is initially active.

\paragraph{\label{sec:discussion}Discussion.}

We have shown that there are neural networks with spike times that vary continuously or even smoothly with network parameters; ordinary spikes only (dis-)appear at the trial end and can be extended to pseudospikes. The networks allow to learn the timings of an arbitrary number of spikes in a continuous fashion with a spike-based gradient.

Perhaps surprisingly, the networks may consist of rather simple, standard QIFs. These are widely used in theoretical neuroscience~\cite{I07,gerstner2014neuronal}, also for supervised learning~\cite{kim18learning,Huh2018,mckennoch2009theta}, and have been implemented in neuromorphic hardware~\cite{basham2009analog,basham2018neuromorphic}. However, the particularity that spikes only (dis-)appear at the trial end has not been noticed and exploited. 
We expect that also further neuron models exhibit spikes with continuous timings, if their voltage slope close to the threshold is guaranteed to be positive.
This includes neuron models that generate spikes by reaching infinite voltage such as hybrid leaky integrate-and-fire neurons with an attached, non-linear spike generation mechanism~\cite{pospischil2011comparison}, the Izhikevich neuron with minor modifications~\cite{I07}, the rapid theta neuron~\cite{Monteforte2011,Engelken2017}, the sine neuron~\cite{Viriyopase2018} and the exponential integrate-and-fire neuron~\cite{gerstner2014neuronal}. It further includes intrinsically oscillating LIFs and anti-leaky integrate-and-fire neurons~\cite{manz2019dynamics}, if the impact of synaptic input currents vanishes at their spike threshold. We also expect that synapses with continuous current rise will be feasible, as well as conductance-based synapses.

On the one hand, our scheme possesses the same advantages as other spike-based gradient descent approaches such as small memory and computational footprints and a clear interpretation as following the exact loss gradient. On the other hand, like standard machine learning schemes it produces no disruptive transitions during learning and no gradient divergences.
This suggests a wide range of applications: When studying biological neural networks, our scheme may be used to learn neurobiologically relevant tasks, to benchmark biological learning, to investigate how the network dynamical solutions may work and to reconstruct synaptic connectivity from experimentally (partially) observed spiking activity. Furthermore, it may be used to train networks in neuromorphic computing (see \cite{Supp} for further discussion). It generally allows to benchmark other learning rules whose underlying mechanisms are less transparent and to (pre-)train networks before converting to a desired neuron type that complicates learning.

The dynamics of spiking and non-spiking neural networks can be chaotic~\cite{sompolinsky1988chaos, Vreeswijk1996,JMT09,MW10, manz2019dynamics} and give rise to exploding gradients~\cite{Supp,goodfellow2016deep,pascanu13rnn,engelken2023lyap}. We therefore restricted our learning examples to at most ten multiples of the membrane time constant. This fits the length of various experimentally observed precisely timed spike patterns~\cite{NHCCB99,Johansson2004firstspikes,gollisch2008rapid,luczak2009spontaneous,havenith2011synchrony,stella22comparing} and the fast processing of certain tasks in neuromorphic computing~\cite{wunderlich2021eventbased,Mostafa2018temporal,Comsa2020snnalpha,Göltz2021neuromorphic,sakemi2023snntemporal}.

Our pseudospikes allow the gradient to \enquote{see} spikes before they appear and to thus add spikes in systematic manner. Pseudospikes affect the pseudospikes of postsynaptic neurons and ultimately of the output neurons. This preserves the gradients of the ordinary spike times and solves, in particular, the dead neuron problem. In a somewhat related approach, silent output neurons were assumed to spike at the trial end~\cite{nowotny2022loss,Kheradpisheh2020s4nn}. Our pseudospikes, however, apply to all neurons and allow to backpropagate errors through silent neurons. The resulting possibility to initialize an entire network with small weights may be important to induce desirable and biologically plausible features such as energy-efficient final connectivity and sparse spiking~\cite{howarth2012updated,Pfeiffer2018}, sparse coding~\cite{olshausen04sparse} and representation learning~\cite{flesch22orthogonal}.  

To conclude, the present study shows something that seemed fundamentally impossible~\cite{roy2019spikebased}: despite the inherent discreteness of spikes, there can be exact non-disruptive, even smooth gradient descent learning in spiking neural networks, including the gradient-based removal and after augmentation also generation of spikes.

\begin{acknowledgments}
We thank Sven Goedeke for helpful comments on the manuscript and the German Federal Ministry of Education and
Research (BMBF) for support via the Bernstein Network
(Bernstein Award 2014, 01GQ1710).
\end{acknowledgments}

\providecommand{\noopsort}[1]{}\providecommand{\singleletter}[1]{#1}%

\end{document}


\title{Smooth Exact Gradient Descent Learning in Spiking Neural Networks\\
\textemdash{} Supplemental Material \textemdash{}}

\author{Christian Klos}
 \email{cklos@uni-bonn.de}
\author{Raoul-Martin Memmesheimer}
 \email{rm.memmesheimer@uni-bonn.de}
\affiliation{Neural Network Dynamics and Computation, Institute of Genetics, University of Bonn, 53115 Bonn, Germany}


\maketitle
\tableofcontents

\section{Materials and methods}
\subsection{Neuron models}

\subsubsection{QIFs with extended coupling}
\label{sec:QIF extended}

We focus in our article on quadratic integrate-and-fire neurons (QIFs)~\cite{latham2000intrinsic,I07,gerstner2014neuronal} that obey the ordinary differential equation
\begin{align}\label{eq:QIF general}
\dot{V}(t)=V(t)(V(t)-1)+I(t).
\end{align}
If $V$ reaches infinity, $V(\tspike^-)=V_\Theta=\infty$, 
an output spike is generated, and the voltage is reset to negative infinity, $V(\tspike^+)=V_\reset=-\infty$. The superscripts $-$ and $+$ denote the limits from the left and right, respectively, which may be interpreted as the times immediately before and after $\tspike$. For small $V$ \cref{eq:QIF general} reduces to the LIF equation \cref{eq:LIF general} with dimensionless membrane time constant $1$. Time is thus measured in multiples of the membrane time constant. Further, we have scaled and shifted the voltage such that without input the QIF has a stable fixed point at the resting potential $V_\rest=0$ and an unstable fixed point at the separatrix potential $V_\sep=1$: For $I(t)=0$ and $V_0=V(0)=V_\rest$ or $V_0=V_\sep$, the right hand side (rhs) of \cref{eq:QIF general} is zero, such that one has a fixed $V(t)=V_0$. If $V_0<0$, the rhs is positive and $V$ increases towards $V_\rest$. Similarly, if $V_\sep>V_0>V_\rest$, the rhs is negative and $V$ decreases towards $V_\rest$. If $V_0>V_\sep$ the rhs is positive, $V(t)$ accelerates towards infinity and a spike is generated. $V_\sep$ thus separates the two classes of trajectories with qualitatively different behavior.

For $I(t)=0$ one can solve \cref{eq:QIF general} by separation of variables. With the initial condition $V(0)=V_0$ the time course  of the voltage reads
\begin{align}\label{eq:V analytic I=0}
    V(t)&=\frac{V_0}{V_0-(V_0-1)\exp(t)}.
\end{align}
\cref{eq:V analytic I=0}'s rhs denominator, $V_0-(V_0-1)\exp(t)$, is at $t=0$ positive (equal to $1$). If $V_0>V_\sep$, it thereafter decreases as the subtrahend $(V_0-1)\exp(t)$ increases with time. The denominator becomes zero when $t$ equals the spike time
\begin{align}\label{eq:t analytic I=0}
    \tspike&=\ln\left(\frac{V_0}{V_0-1}\right),
\end{align} such that $V(\tspike)=\infty$. \tspike depends smoothly on $V_0$ and if $V_0$ tends to $V_\sep$, \tspike tends to infinity.

The input current $I(t)$ consists of contributions due to spikes arriving from other neurons in the considered network. Additionally, there may be a constant input current component $I_0$, which covers average input from further neurons that are not explicitly modeled.
To model temporally extended synaptic coupling, we implement standard current-based exponentially decaying synapses~\cite{vogels2005neural,B06a,memmesheimer2014learning}. Specifically, at a spike arrival time $t_{i}$ of a spike from neuron $i$, $I(t)$ increases about the strength $w_i$ of the synapse from neuron $i$. Between spike arrivals, the current decays exponentially with time constant $\taus$. $I(t)$ thus obeys
\begin{align}\label{eq:I}
\taus \dot{I}(t)=-(I(t)-I_0)+ \taus \sum_i w_i \sum_{t_i}\delta(t-t_i),
\end{align}
with the Dirac delta distribution $\delta$. We focus on neurons with $I_0=0$ in our article.

\subsubsection{A closed-form solution}\label{sec:QIF analytical solution}
Interestingly, \cref{eq:QIF general,eq:I} have a closed-form solution between spikes, if $\taus=1/2$ and $I_0=0$ (in general the solution involves Bessel functions \cite{mathematica}): To obtain it, we first shift the time origin to the beginning of the period of interest. Since the period of interest extends only to the next input or output spike, it does not contain spike arrivals and all synaptic input currents decay exponentially. Since, furthermore, all synaptic decay time constants are identical, this implies that we can gather the currents into a single exponentially decaying one: \cref{eq:I} with $I_0=0$ yields
\begin{align}
I(t)=\sum_{i}w_i\sum_{t_i\leq 0}e^{-\frac{t-t_i}{\tau_s}}=\Big(\sum_{i}w_i\sum_{t_i\leq 0}e^{-\frac{0-t_i}{\tau_s}}\Big)e^{-\frac{t-0}{\tau_s}}=I(0)e^{-\frac{t-0}{\tau_s}}=w e^{-2t},
\end{align}
 where we have called the current strength at the time origin $w=I(0)$ and inserted $\tau_s=1/2$.
Using this in \cref{eq:QIF general} gives
\begin{align}
    \label{eq:QIF between spikes}
    \dot{V}(t)=V(t)^2-V(t)+w e^{-2t}.
\end{align}
The simple substitution $V(t)=e^{-t}u(t)$ leads to a differential equation for $u(t)$ where the variables separate,
\begin{align}
    \dot{u}(t)=(u^2(t)+w)e^{-t},
\end{align}
\cite{kamke1977differentialgleichungen} (part C, Eq.~(I$\cdot$55)). The solution of \cref{eq:QIF between spikes} with $V(0)=V_0$ is thus
\begin{align}\label{eq:QIF analytic solution}
    V (t)=
    \begin{cases}
    \frac{V_0}{V_0-(V_0-1)\exp(t)},&\text{if }w=0,\\
        \sqrt{w}e^{-t}\tan\left(\arctan\left(\frac{V_0}{\sqrt{w}}\right)+\sqrt{w}\left(1-e^{-t}\right)\right), &\text{if }w>0,\\
        \sgn(V_0)\sqrt{-w}e^{-t}, &\text{if $w<0$ and $|V_0|=\sqrt{-w}$},\\
        \sqrt{-w}e^{-t}\coth\left(\arcoth\left(\frac{V_0}{\sqrt{-w}}\right)-\sqrt{-w}\left(1-e^{-t}\right)\right), &\text{if $w<0$ and $|V_0|>\sqrt{-w}$},\\
        \sqrt{-w}e^{-t}\tanh\left(\artanh\left(\frac{V_0}{\sqrt{-w}}\right)-\sqrt{-w}\left(1-e^{-t}\right)\right), &\text{if $w<0$ and $|V_0|<\sqrt{-w}$}.
    \end{cases}
\end{align}

This solution yields closed-form conditions for the generation of output spikes and even closed-form expressions for the spike times. The case $w=0$ is discussed in the previous paragraph. A spike is generated if $V_0>V_{\sep}=1$; \cref{eq:t analytic I=0} provides the spike time.
If $w>0$, we have a spike under the condition that $\sqrt{w}+\arctan\left(\frac{V_0}{\sqrt{w}}\right)>\frac{\pi}{2}$: The argument of $\tan$ in the second line of \cref{eq:QIF analytic solution} is initially smaller than $\pi/2$ because $\arctan\left(\frac{V_0}{\sqrt{w}}\right)<\frac{\pi}{2}$ and the second summand is zero. The condition ensures that for time tending to infinity the argument exceeds $\pi/2$, since $e^{-t}$ tends to zero. Therefore for some finite spike time $\tspike$, the argument reaches $\pi/2$ from below and $\tan$ and $V(t)$ tend to positive infinity when $\tspike$ is approached. Setting the argument equal to $\pi/2$ yields
\begin{align}
    \tspike=-\ln\left(1-\frac{\pi}{2\sqrt{w}}+\frac{1}{\sqrt{w}}\arctan\left(\frac{V_0}{\sqrt{w}}\right)\right).\label{eq:QIF ext tspike}
\end{align}
For $w<0$, there is no spike generation if $|V_0|\leq\sqrt{-w}$, because the solutions are bounded by $\sqrt{-w}$. If $V_0>\sqrt{-w}$ there a spike is generated under the condition that  $\arcoth\left(\frac{V_0}{\sqrt{-w}}\right)-\sqrt{-w}<0$ holds: the argument of $\coth$ in the third line of \cref{eq:QIF analytic solution} is initially positive, since $\arcoth$ is positive for arguments larger than $1$. The condition ensures that for time to infinity the argument becomes smaller than zero, since $e^{-t}$ tends to zero. Therefore the argument reaches zero at a finite time from the positive side such that $\coth$ and $V(t)$ tend to positive infinity. This happens at
\begin{align}
    \tspike=-\ln\left(1-\frac{1}{\sqrt{-w}}\arcoth\left(\frac{V_0}{\sqrt{-w}}\right)\right).\label{eq:QIF ext tspike w-}
\end{align}

\subsubsection{Phase representation}
For the second type of pseudospike times (\cref{sec:QIF pseudo}) and for our analytical considerations (\cref{sec:QIF ordinary proof}), we transform the voltage of QIFs with extended coupling to an angle variable. In other words, we transform the QIF to a $\theta$-neuron~\cite{EK86,latham2000intrinsic,I07,gerstner2014neuronal}. The transformation is smooth, i.e.~continuously differentiable, and bijective, except at spiketimes, where $V$ becomes infinitely large and is reset.
Concretely, we use
\begin{align}\label{eq:Phi}
    \phi=\Phi(V)=\frac{1}{\pi}\arctan\left(\frac{V}{\pi}\right)+\frac{1}{2},
\end{align}
such that the threshold and reset of $\phi$ are $\phi_\Theta=1$ and $\phi_\reset=0$.
Identifying the phases of threshold and reset with each other lets the $\phi$-dynamics take place on a circle, $S^1$. They obey the differential equation
\begin{align}\label{eq:phi dot}
    \dot{\phi}(t)=\frac{1}{\pi}\frac{\dot{V}(t)/\pi}{1+(V(t)/\pi)^2}=\cos(\pi\phi(t))\left(\cos(\pi\phi(t))+\frac{1}{\pi}\sin(\pi\phi(t))\right)+\frac{1}{\pi^2}\sin^2(\pi\phi(t))I(t),
\end{align}
where we used \cref{eq:Phi,eq:QIF general} and $V=\Phi^{-1}(\phi)=-\pi\cot(\pi\phi)$. The point $\phi=1$, which is the same as $\phi=0$, is not particularly special anymore, as the right hand side of the differential equation is infinitely often continuously differentiable there. $\phi$'s temporal derivative at this point equals $1$, independent of $I$.

\subsubsection{QIFs with infinitesimally short coupling}
\label{sec:QIF inf short}

Furthermore, we consider QIFs with input currents of infinitesimally short extent~\cite{T88a,T88b,Brunel00,B06a,MT06A,Memmesheimer2010}. These induce a jump-like response in the voltage upon input arrival. Specifically, at a spike arrival from neuron $i$, $V(t)$ increases by the synaptic strength $w_i$. $V(t)$ and $I(t)$ are thus determined by
\begin{align}
\taum \dot{V}(t) &= V(t)(V(t)-1) + I(t),\\
I(t) &= I_0 + \taum\sum_i w_i \sum_{t_i}\delta(t-t_i).\label{eq:input current inf short input}
\end{align}
Here, $\taum$ is the membrane time constant, $I_0$ is the constant input current component and, as before, the voltage threshold is $V_\Theta=\infty$ and the reset potential $V_\reset=-\infty$. 

In our simulations, we always use a suprathreshold constant input current, i.e.~$I_0>1/4$, which ensures that $\dot{V}(t)$ is positive if there is no further input. Hence, the neurons are intrinsically oscillating.
Their dynamics between spikes is simplified: they have no fixed points anymore and the voltage is always monotonously increasing.
We transform the QIF to a $\Theta$-neuron, using the transformation
\begin{align}
    \phi = \Phi(V) &= \frac{\taum}{\sqrt{I_0-\frac{1}{4}}}\left(\arctan\left(\frac{V-\frac{1}{2}}{\sqrt{I_0-\frac{1}{4}}}\right)+\frac{\pi}{2}\right),\\
    V = \Phi^{-1}(\phi) &= \sqrt{I_0-\frac{1}{4}}\tan\left(\sqrt{I_0-\frac{1}{4}}\frac{\phi}{\taum}-\frac{\pi}{2}\right)+\frac{1}{2}.
\end{align}
The threshold and reset of $\phi$ are then given by $\phi_\Theta=\Phi(\infty)=\taum\pi/\sqrt{I_0-\frac{1}{4}}$ and $\phi_\reset=\Phi(-\infty)=0$, respectively. We choose a slightly different transformation than before (cf.~\cref{eq:Phi}), because it results in a constant phase velocity between spikes,
\begin{align}\label{eq:Theta short}
    \dot{\phi}(t) = 1.
\end{align}
The closed-form solution of \cref{eq:Theta short} between spikes and with $\phi(0)=\phi_0$ is simply $\phi(t)=\phi_0+t$. If there are no spike arrivals, the next spike thus happens at $\tspike=\phi_\Theta-\phi_0$. Such simple expressions are convenient for event-based simulations. At a spike arrival from neuron $i$ at $t_i$, $\phi$ changes according to the transfer function or phase transition curve $H_{w}(\phi)$ \cite{MS90,MT06B,Viriyopase2018}. Concretely,
\begin{align}
    \phi(t_i^+) = H_{w_i}(\phi(t_i^-))=\Phi\left(\Phi^{-1}(\phi(t_i^-)) + w_i\right).
\end{align}

\subsubsection{LIFs with extended coupling}\label{sec:LIF}
For comparison purposes, we also consider LIFs with extended coupling:
\begin{align}\label{eq:LIF general}
    \dot{V}(t)&=-V(t)+I(t),\\
    \taus \dot{I}(t)&=-(I(t)-I_0)+ \taus \sum_i w_i \sum_{t_i}\delta(t-t_i),
\end{align}
where $I_0$ is the constant input current component and $i$ indexes the presynaptic neurons with corresponding synaptic weights $w_i$ and spike times $t_i$. Time, including the synaptic time constant $\taus$, is measured in multiples of the membrane time constant $\taum$ and the voltage has been shifted and scaled such that the resting potential is at $V_\rest=0$ and the threshold at $V_\Theta=1$. Directly after reaching the threshold, the voltage is reset to $V_\reset=V_\rest$.

Assuming $I_0=0$ and $\taus\neq\taum$, the closed-form solution of \cref{eq:LIF general} with $V(0)=V_0$ and $I(t)=we^{-t/\taus}$, i.e. between spikes, is given by
\begin{align}\label{eq:LIF solution}
    V(t) = V_0 e^{-t} + w\frac{\taus}{1-\taus}(e^{-t}-e^{-\frac{t}{\taus}}).
\end{align}
If $\taus=1/2$, the rhs of \cref{eq:LIF solution} is quadratic in $e^{-t}$, which allows to compute the time of the next spike, in case there is one, in closed form~\cite{haene2009liftau}. Specifically, the threshold crossing happens at
\begin{align}
    \tspike = -\ln\left(\frac{1}{2w}\left(V_0 + w + \sqrt{(V_0 + w)^2-4wV_\Theta}\right)\right).
\end{align}
Here we assumed that the argument of the logarithm lies between 0 and 1, which ensures that $V(t)$ reaches $V_\Theta$.

\subsection{Pseudospikes}
\label{sec:QIF pseudo}

\subsubsection{First type of pseudospikes for QIFs with extended coupling}
\label{sec:QIF pseudo1}

In this section, we explain the first type of pseudodynamics and pseudspikes for QIFs with extended coupling, cf.~\cref{sec:QIF extended}. For the pseudodynamics we assume that the neurons behave like freely evolving QIFs with an added, constant drive after the trial end. Specifically, we define them to be
\begin{align}\label{eq:QIF pseudo1}
    \dot{V}_\pseudo(t) = \Vpseudo(t) (\Vpseudo(t)-1) + \frac{1}{4} + g(\Ipseudo)
\end{align}
with initial condition $\Vpseudo(T)=V(T)$, where $T$ is the trial length. $\Ipseudo$ is a modified version of the input current at the trial end $I(T)$, see below, and $g(I)=\alpha\log(1+\exp(I / \alpha))$ with a free parameter $\alpha>0$. Choosing the pseudodynamics to also be quadratic ensures the smooth transition of ordinary spike times to pseudospike times (see \cref{sec:QIF pseudo1 proof1}). The added, suprathreshold drive $I_0=\frac{1}{4} + g(\Ipseudo)$ ensures that the pseudodynamics are oscillatory ($g(\Ipseudo)$ is positive), such that pseudospikes are generated. 

One can transform the voltage of the pseudodynamics with the same transformation as in \cref{sec:QIF inf short} to an angle variable, 
\begin{align}
\label{eq:phi pseudo1}
    \phi_\pseudo = \Phi_{\Ipseudo}(V) = \frac{1}{\sqrt{g(\Ipseudo)}} \left(\arctan\left(\frac{V-1/2}{\sqrt{g(\Ipseudo)}}\right)+\frac{\pi}{2}\right).
\end{align}
The threshold and reset of $\phi_\pseudo$ are then given by $\phi_{\Theta,\Ipseudo}=\pi/\sqrt{g(\Ipseudo)}$ and $\phi_\reset=0$, respectively. \cref{eq:QIF pseudo1} transforms to $\dot{\phi}_\pseudo = 1$, making the computability of the pseudospike time in closed form obvious. Specifically, the general expression for the time of the $k$th spike, in case it is a pseudospike, is
\begin{align}\label{eq:QIF pseudo1 tpseudo}
    \tpseudo &= T + (k - n_\trial) \phi_{\Theta,\Ipseudo} - \Phi_{\Ipseudo}(V(T)),
\end{align}
where $n_\trial$ is the number of ordinary spikes. The factor $(k - n_\trial)$ ensures continuity of spiketimes whenever the current or a previous spike time crosses the trial end (see \cref{sec:QIF pseudo1 proof2} for details). For example, if an ordinary spike becomes a pseudospike, $-\Phi_{\Ipseudo}(V(T))$ jumps by $-\phi_{\Theta,\Ipseudo}$ since the reset crosses $T$. This is canceled by the simultaneous jump of $(k - n_\trial)\phi_{\Theta,\Ipseudo}$ by $\phi_{\Theta,\Ipseudo}$, since $n_\trial$ decreases by one. The spiketimes $\tpseudo$ thus change continuously.

To ensure generically non-zero gradients, the pseudospike times should be affected by other neurons even if they are not generating ordinary spikes. During the trial, a presynaptic spike leads to a jump of the input current about the synaptic weight. Inspired by this, we here assume that presynaptic neurons affect the constant input current $I_0$ by a fraction of the synaptic weight. Specifically, we set
\begin{align}\label{eq:QIF pseudo1 Ipseudo}
    \Ipseudo &= I(T) + \sum_{j} w_j \frac{\Phi_{I_{\pseudo,j}}(V_j(T))}{\phi_{\Theta,I_{\pseudo,j}}},
\end{align}
where $j$ indexes the presynaptic neurons. Thus, for each neuron $j$ a fraction of its synaptic weight $w_j$ is added to the input current at the trial end $I(T)$. This fraction depends on how close neuron $j$ is to producing a spike at the trial end, reaching one when the neuron reaches the threshold there. The additional input ensures that errors can be backpropagated through silent neurons and guarantees continuity of \Ipseudo in case a presynaptic spike from neuron $j$ crosses the trial end: then $I(T)$ jumps by $w_j$, which is canceled because $V_j(T)$ jumps from $\infty$ to $-\infty$, which induces a jump in $\Phi_{I_{\pseudo,j}}(V_j(T))$ by $-\phi_{\Theta,I_{\pseudo,j}}$ (see \cref{sec:QIF pseudo1 proof3} for details). Finally, one needs to specify the initial values of \Ipseudo and the order in which \cref{eq:QIF pseudo1 Ipseudo} is evaluated given the neural network architecture. Specifically, in a feedforward network, we set $\Ipseudo=I(T)$ for the neurons in the first layer and then use \cref{eq:QIF pseudo1 Ipseudo} to sequentially compute \Ipseudo for the other layers. In a recurrent network, we first set $\Ipseudo=0$ for all neurons and then use \cref{eq:QIF pseudo1 Ipseudo} to compute the final values of \Ipseudo. These choices ensure the validity of the pseudospike properties listed in the main text. 

The scaling factor in \cref{eq:QIF pseudo1 Ipseudo} can be rewritten as
\begin{align}
    r_j  &= \frac{\Phi_{I_{\pseudo,j}}(V_j(T))}{\phi_{\Theta,I_{\pseudo,j}}}
    = \frac{t^{\max}_{\pseudo,j}-t_{\pseudo,j}}{t^{\max}_{\pseudo,j}-T}.\label{eq:QIF pseudo1 r}
\end{align}
Here, $t_{\pseudo,j}$ is the first pseudospike time of neuron $j$ and 
\begin{align}
    t^{\max}_{\pseudo,j} &= T + \phi_{\Theta,I_{\pseudo,j}}\label{eq:QIF pseudo1 tmax}
\end{align}
is its latest possible timing, which occurs for $V_j(T)\rightarrow -\infty$. This shows that neurons with earlier first pseudospike have a stronger influence on the pseudospike times of their postsynaptic partners. Furthermore, \cref{eq:QIF pseudo1 r,eq:phi pseudo1,eq:QIF pseudo1 Ipseudo} show that $r_i$ may be expressed as
\begin{align}
    r_i = f_i\big(\sum_{j} w_j r_j\big).
\label{eq:QIF pseudo1 ri}
\end{align}
Thus, we can compute the pseudospike times like the states in a network of rate neurons that is run for one time step. Comparing \cref{eq:QIF pseudo1 ri} and \cref{eq:QIF pseudo1 r} yields the activation function
\begin{align}
    f_i(x)=\left.\frac{\Phi_{I_{\pseudo,i}}(V_i(T))}{\phi_{\Theta,I_{\pseudo,i}}}\right|_{\sum_{j} w_j r_j=x}=\frac{1}{\pi}\arctan\left(\frac{V_i(T)-1/2}{\sqrt{g(I_i(T)+x)}}\right)+\frac{1}{2}.
\end{align}
In contrast to common networks of rate neurons, the activation function generally changes in each learning step, as it depends on $V_i(T)$ and $I_i(T)$.

\begin{figure}[h!]
    \centering
    \includegraphics{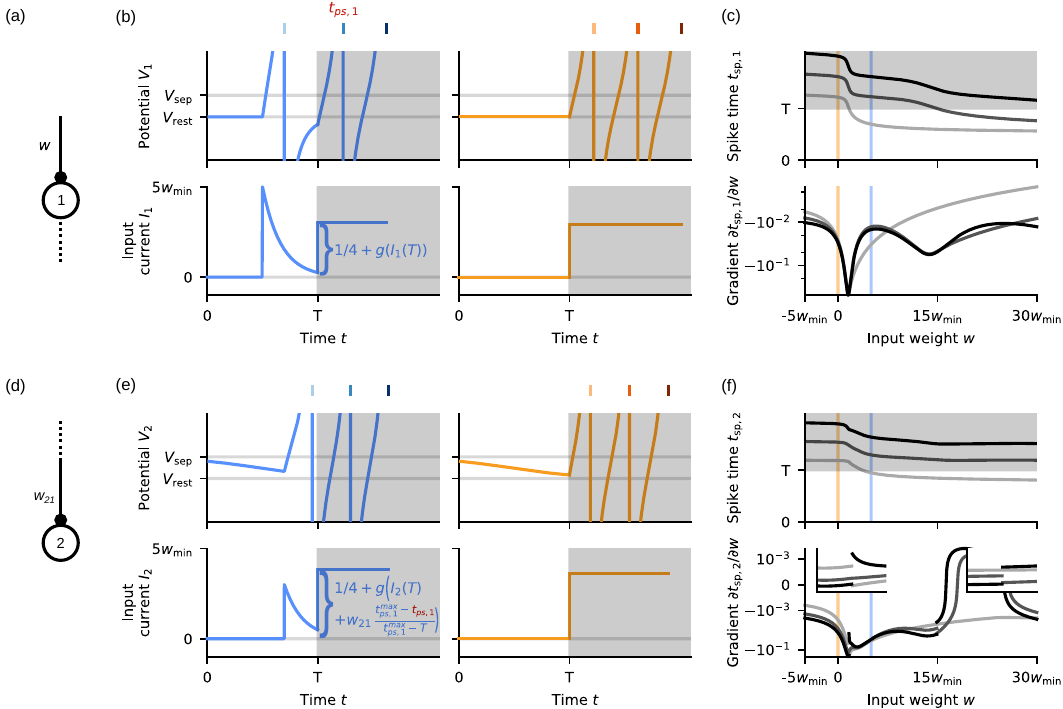}
    \caption{\label{fig:Pseudospikes type 1} First type of pseudodynamics and pseudospikes. The figure shows in panels (a-c) results of simulations of a single neuron (neuron 1) that receives a single input spike; another neuron (neuron 2) is connected to neuron 1 and receives its output spikes (d-f). (a) Schematics of neuron 1, highlighting that it has a single input connection with weight $w$ and a single output connection. (b) Ordinary and pseudodynamics of neuron 1 for two different weight values.
    Left, blue traces: $w=5w_{\min}$, where $w_{\min}$ is the weight at which an ordinary spike appears at infinity (cf.~main text Fig.~1). During the ordinary dynamics (white background), the input current (lower panel) due to the input spike is strong enough to induce an ordinary spike (upper panel, light blue vertical tick). Right, orange traces: $w=0$, the neuron does not generate ordinary spikes.
    During the pseudodynamics (gray background) the input current is set to a constant, suprathreshold value. This value depends on the input current at the trial end, $I_1(T)$, and is therefore different for the blue and orange traces. The pseudodynamics start at $V_1(T)$ and generate pseudospikes. The pseudodynamics continue until the desired number of spikes is generated, which we here assume to be three. (c) Times of the three spikes of neuron 1 and their derivatives with respect to the input weight, as a function of the input weight. The spike times and their derivatives are continuous, which means that gradient descent can be used to smoothly shift spike times into the trial. Vertical lines correspond to similarly colored examples shown in (b). (d) Schematics of neuron 2, highlighting that it receives a connection with weight $w_{21}$ from neuron 1. (e) Same as (b) but for neuron 2. The value of the input drive during the pseudodynamics depends on $I_2(T)$ and on the first pseudospike time of the presynaptic neuron 1 (dependency highlighted in red). Therefore it is different for the blue and orange traces. (f) Same as (c) but for neuron 2. The spike times are continuous and mostly smooth. Discontinuities of the derivatives of pseudospike times (insets) appear when a spike of the presynaptic neuron 1 crosses the trial end $T$. Gradient descent can be used to shift spike times into the trial even if neither neuron 2 nor neuron 1 spike during the trial.}
\end{figure}

In \cref{sec:QIF pseudo1 proof}, we show the continuity and mostly smoothness of the here defined pseudospike times. \cref{fig:Pseudospikes type 1} illustrates the computation of pseudospike times and the continuous and mostly smooth dependence of spike times on network parameters. 

\subsubsection{Second type of pseudospikes for QIFs with extended coupling}
\label{sec:QIF pseudo2}

In the following, we explain the second type of pseudodynamics and pseudspikes. They are based on extending the dynamics behind the trial end, first according to the same differential equations that also govern the ordinary dynamics, thereafter according to simplified dynamics. For consistency with our explanation of the first type of pseudodynamics and pseudspikes, it makes most sense to call all dynamics and spikes within the trial duration $[0,T]$ ordinary dynamics and ordinary spikes. We use this convention in the main text; it implies that pseudospikes only influence pseudospikes, property (iii) in the main text. The pseudodynamics then consist of two parts, a first part obeying the same dynamical equations as the ordinary dynamics and a second part obeying modified dynamics. To simplify the subsequent explanations, we choose a different convention in the current section: We denote all dynamics that obey the ordinary dynamics equations by ordinary dynamics, even if they extend beyond $T$. Further, we call only the modified dynamics pseudodynamics.

The basic ideas behind the construction of the second type of pseudodynamics and pseudspikes may then be gathered as follows: (i) The ordinary neuronal dynamics guarantee smoothness of spikes, so we use it during the time when inputs arrive. (ii) (Active) Pseudospikes depend only on the phase at the end of the ordinary dynamics. (iii) If an ordinary spike disappears or appears a corresponding pseudospike appears or disappears at the beginning of the ensuing period of pseudodynamics. (iv) (Active) Pseudospikes that change to ordinary spikes are immediately replaced, such that there is always exactly one pseudospike per neuron. (v) The precise functional dependence of (active) pseudospike times on the phase at the end of the trial is such that spike times change smoothly with the network parameters also for special events like pseudospikes becoming ordinary ones.

\begin{figure}[h!]
    \centering
    \includegraphics{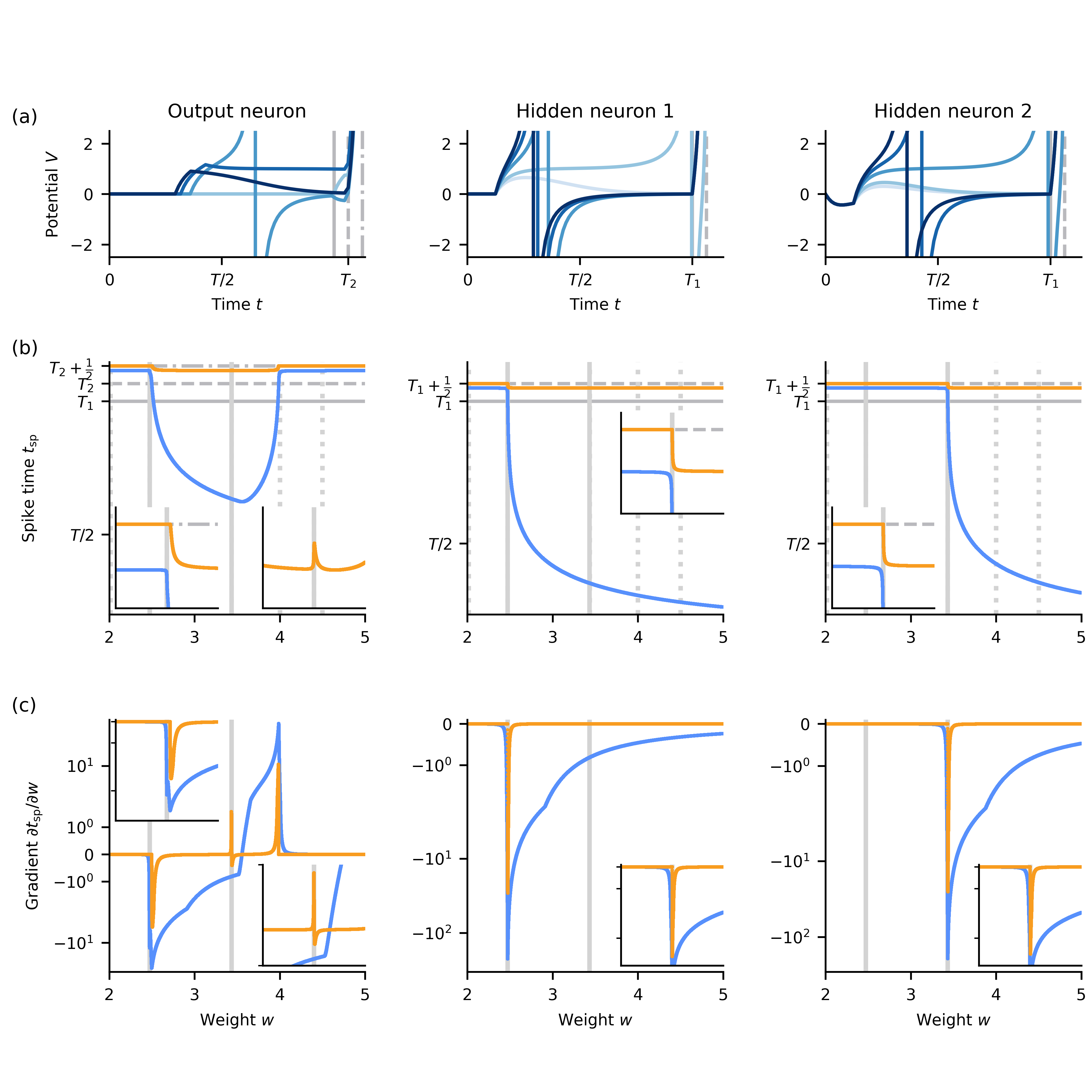}
    \caption{\label{fig:Pseudospikes type 2} Second type of pseudodynamics and pseudospikes. The figure shows the results of simulations in a basic two-layer network with two hidden neurons and one output neuron. There is one input at the beginning of the trial, which inhibits hidden neuron $2$, and one input a bit later, which excites both hidden neurons by $w$. Hidden neuron $1$ excites the output neuron, hidden neuron $2$ inhibits it. (a) Voltage traces of the output and the two hidden neurons for increasing $w$
    plotted in increasing color intensity. The pseudodynamics with $d=2$ take place within $(T_1,T_1+1/d]$ and $(T_2,T_2+1/d]$ in the hidden and the output neurons, respectively. 
    Solid, dashed and dashed-dotted vertical gray lines indicate $T_1$, $T_1+1/d=T_2$ and $T_2+1/d$, respectively. 
    (b) Spike times as a function of $w$ (blue, orange: first, second spike time of the different neurons). For increasing $w$ there are transitions from an active pseudospike to an ordinary spike and simultaneously from an inactive to an active pseudospike, first in hidden neuron $1$ then in $2$. The insets show closeups of the curves around the corresponding weight values ($w\approx 2.47, 3.43$, solid gray vertical lines; spike time axis magnifications differ). The spiking of the hidden neurons and its temporal change trigger similar transitions in the output neuron. Dotted and solid vertical lines indicate weight values of traces displayed in (a). (c) like (b) for the gradient of the spike times with respect to $w$. The curves in (b,c) are continuous, because the spike times are smooth in $w$. This holds in particular at the transitions between inactive and active pseudospikes and between active pseudospikes and ordinary spikes.}
\end{figure}

In detail, we consider a feedforward network of $L$ layers. The trial and thus the input spike trains last until $T$. The ordinary dynamics of the neurons in each layer beyond the first hidden layer are increasingly extended: in layer $l={1,...,L}$ they last until
\begin{align}\label{eq:Tl pseudo 2}
    T_l&=T+\frac{l-1}{d},
\end{align}
i.e.~if we go up one layer, the ordinary dynamics last a fraction $1/d$ of the membrane time constant longer. We assume  $d>1$, which ensures the smoothness of spike times. After the ordinary dynamics, each neuron i in layer $l$ generates pseudodynamics that lead to one pseudospike time 
\begin{align}\label{eq:QIF pseudo2}
\tpseudo&=T_l+\frac{1}{d}-\frac{1}{d}\phi(T_l)^d,
\end{align}
where $\phi(T_l)$ is the phase \cref{eq:Phi} at the end of the ordinary dynamics. If $\phi(T_l)=1$, which is the same state as $\phi(T_l)=0$, the value $0$ is inserted into \cref{eq:QIF pseudo2}, such that $\tpseudo$ lies in the half open interval $(T_l,T_l+\tfrac{1}{d}]$ directly ensuing the period of ordinary dynamics.
The pseudospike times from layer $l-1$ thus arrive at the neurons of layer $l$ towards the end of their  ordinary dynamics. A pseudospike time $\tpseudo$ in a neuron of layer $l$ may be interpreted as resulting from completely externally driven pseudodynamics $\phi_\pseudo$ beyond $T_l$. The continuous matching $\phi_\pseudo(T_l^+)=\phi(T_l)$ to the preceding dynamics and the spike time condition $\phi_\pseudo(\tpseudo)=1$ imply that they can be specified as
\begin{align}
\phi_\pseudo(t)=\phi_\pseudo(T_l^+)+
1-(1-d(t-T_l))^\frac{1}{d},
\end{align}
such that they obey the differential equation
\begin{align}
\dot{\phi}_\pseudo(t)=(1-d(t-T_l))^{\frac{1}{d}-1}.
\end{align}
Using $\Phi^{-1}$ (cf.~\cref{eq:Phi}), they can be transformed into voltage pseudodynamics, as displayed in \cref{fig:Pseudospikes type 2}a. Pseudodynamics with $d=1$ linearly extrapolates the phase $\phi(T_l)$ to the threshold with slope one, such that the pseudospike happens at $\tpseudo=T_l+1-\phi(T_l)$.

If the network parameters change, pseudospikes become ordinary ones and vice versa. The related spiketimes change smoothly. For example, if a pseudospike of a neuron in layer $l$ tends to $T_l$, $\phi(T_l)$ tends to $1$, such that the ordinary spike appears at $T_l$ exactly at the parameter value at which the pseudospike would reach $T_l$ (and vanishes). The spiketime initially related to the pseudospike and then to the ordinary spike thus changes continuously. We assume that all pseudospikes that will be needed in the considered parameter range are held inactive but available at $T_l+\tfrac{1}{d}$. This may be important to construct a smooth cost function, because output layer spikes that are desired but not yet present as active pseudospikes can be included in it. (An alternative assumption compatible with our scheme is that a new pseudospike emerges if the current one becomes an ordinary spike.)

\cref{fig:Pseudospikes type 2}b,c illustrates the smooth dependence of the spiketimes on the network parameters in presence of pseudospikes. One can prove that it holds also at the transitions between inactive pseudospikes, active pseudospikes and ordinary spikes using methods similar to those of \cref{sec:QIF ordinary proof}.

\subsubsection{Pseudospikes for QIFs with infinitesimally short coupling}
\label{sec:QIF inf short pseudo}

For the pseudospikes of QIFs with infinitesimally short coupling (\cref{sec:QIF inf short}), we take a similar approach as for the first type of pseudospikes of QIFs with extended coupling (\cref{sec:QIF pseudo1}). This ensures that the pseudospike times are continuous. Specifically, we define the pseudodynamics to be 
\begin{align}\label{eq:QIF inf short pseudo}
    \taum\dot{V}_\pseudo(t) = \Vpseudo(t) (\Vpseudo(t)-1) + I_0,
\end{align}
where $I_0$ has the same value as in \cref{eq:input current inf short input}. In other words, the neurons continue to evolve as during the trial, but without interactions.

Similar to \cref{sec:QIF pseudo1}, we assume that neurons interact at the trial end with each other in the same way as during the trial but with scaled connection weights. Therefore, we set the initial condition for the pseudodynamics to
\begin{align}
\label{eq:Vpseudo infinitesimal short}
    \Vpseudo(T) = V(T) + \sum_j w_j \frac{\Phi(V_{\pseudo,j}(T))}{\phi_\Theta}
\end{align}
where $j$ indexes the presynaptic neurons and $\Phi(V)$ as well as $\phi_\Theta$ are defined as in \cref{sec:QIF inf short}.
Hence, the time of the $k$th spike, in case it is a pseudospike, is
\begin{align}
    \tpseudo = T + (k-n_\trial) \phi_\Theta - H_{\sum_j w_j \frac{\Phi(V_{\pseudo,j}(T))}{\phi_\Theta}}(\Phi(V(T)),
\end{align}
where $n_\trial$ is the number of ordinary spikes and $H_w(\phi)$ is defined as in \cref{sec:QIF inf short}. 
Similar to \cref{eq:QIF pseudo1 Ipseudo,eq:phi pseudo1}, \cref{eq:Vpseudo infinitesimal short} implies that we can obtain the pseudospike times from the states $V_i(T)$ at the trial end and variables $r_i$ that are computed like the states of a network of rate neurons,
\begin{align}
    r_i  &= \frac{\Phi(V_{\pseudo,i}(T))}{\phi_\Theta}= \frac{\Phi\left( V_i(T) + \sum_j w_j \frac{\Phi(V_{\pseudo,j}(T))}{\phi_\Theta}\right)}{\phi_\Theta}\\
    &= \frac{\Phi\left( V_i(T) + \sum_j w_j r_j\right)}{\phi_\Theta}
    =f_i\big(\sum_j w_j r_j\big).
\end{align}

\subsection{Simulation details}

We mostly use exact, event-based simulations~\cite{brette2007snnsim}, where one iterates over spikes using the closed-form solutions for the evolution of the dynamical variables and upcoming spike times, see \cref{sec:QIF analytical solution,sec:QIF inf short,sec:LIF}. In each iteration, at first the neuron that spikes next as well as the time of the next spike is determined. Second, the state of all neurons is evolved until the next spike time. Third, the state of the neurons postsynaptic to the spiking neuron is updated based on the synaptic mechanism. Finally, the state of the spiking neuron is reset. For numerical reasons, some minor approximations are necessary if the absolute value of the membrane potential gets very large (see next paragraph). In \cref{fig:Pseudospikes type 2,fig:tsp(w),fig:tsp(ti),fig:tsp(wi),fig:GradientStatistics}, we use time step based simulations, employing a standard ordinary differential equation solver between input and output spikes, with event detection to detect threshold crossings.

We simulate QIFs with extended coupling mostly in $V$-space. For the event-based simulations, we neglect the effect of an incoming spike on the next spike time of a neuron, if the spike time is less than $\varepsilon$ away, where $\varepsilon=10^{-6}$. Further, we do not update $V$ if it is greater than $1/\varepsilon$ anymore and after spike generation at positive infinity, we reset $V$ to $-1/\varepsilon$. For numerical purposes these values are sufficiently close to $\pm\infty$. In \cref{fig:GradientStatistics}, we employ time-step-based voltage and current simulations with a threshold of $10^5$ and a reset of $-10^5$. \cref{fig:Pseudospikes type 2,fig:tsp(w),fig:tsp(ti),fig:tsp(wi)} use time-step-based phase and current simulations with threshold $1$ and reset $0$.

We simulate QIFs with infinitesimally short coupling in $\phi$-space using event-based simulations. We neglect the effect of an incoming spike on $\phi$ and thus also the next spike time, if $\phi$ is very close to the threshold, $\phi>\Theta - \varepsilon$, or very close to the reset $\phi<\varepsilon$, where $\varepsilon=10^{-6}$.

We simulate LIFs with extended coupling in $V$-space using event-based simulations. This is possible since we set the synaptic time constant to half of the membrane time constant. In this case, a closed-form solution of the threshold crossing time is available, see \cref{sec:LIF}.

We use Python for all our simulations and analysis. For the event-based simulations and the automatic differentiation, we use JAX~\cite{bradbury2018jax}. For the time step-based simulations, we use NumPy~\cite{harris2020numpy}
and SciPy~\cite{virtanen2020scipy}. 
Further, we use PyTorch~\cite{paszke2019pytorch} for data loading, Optax~\cite{deepmind2020optax} for optimization and Ray~\cite{moritz2018ray} for hyperparameter search. For plotting, we use Matplotlib~\cite{hunter2007matplotlib} with colorblind-friendly colors~\cite{petroff2021colorblind}. All simulations were run on a local workstation with consumer-grade CPU (AMD Ryzen 1800X) and GPU (NVIDIA GeForce RTX 3090). Code to reproduce the main results is publicly available~\cite{klos2024code}.

\subsection{Spike time arc length}

In some of our figures, we plot the evolution of spike times during learning as a function of the arc length of the spike time trajectories. At trial $n$, this is the cumulative, absolute change of all learned spike times until $n$:
\begin{align}
    L_t(n) = 
    \begin{cases}
        0, &\text{if }n=0,\\
        \sum\limits_{l=1}^{n} \sum\limits_{i=1}^{N_\target}\sum\limits_{k_i=1}^{N_{\target,i}} |t_{k_i}^l -  t_{k_i}^{l-1}|, &\text{else},
    \end{cases}
\end{align}
where $l$ indexes the trial, $i$ indexes the $N_\target$ neurons whose spike times are learned, $k_i$ indexes the $N_{\target,i}$ learned spike times of neuron $i$ and and $t_{k_i}^l$ is the time of spike $k_i$ at trial $l$. In \cref{fig:fig3s1,fig:fig3s2}, we additionally smooth the spike times with a rectangular kernel of length 11 trials before computing the spike time arc length to reduce the effect of oscillations on $L_t(n)$.

\newpage

\section{Non-disruptive (dis-)appearance of spikes and smooth spike timing in QIFs with extended coupling}\label{sec:QIF ordinary proof}

The following section shows that in QIFs with temporally extended coupling the output spike times depend smoothly on the input spike times and the input weights and that spikes can only (dis-)appear at the trial end. We note that for clarity of the arguments, in this section we allow this trial end to be at temporal infinity. We conduct the proof for arbitrary synaptic time constant; for simplicity, we assume that there is no constant input current. The proof uses well-known facts from analysis and the theory of differential equations. We sketch it in the next subsection, \cref{sec:proof overview}.
Thereafter we detail it in five subsections that build on each other: \cref{sec:Smooth dependence on the initial conditions} shows smooth dependence of later states and spike times on the initial states. The initial state of the input current may be interpreted as the weight strength of a single input that arrives at the initialization time. \cref{sec:Smooth dependence on the input weights} generalizes this result by separating time into intervals in each of which one input arrives at the beginning. \cref{sec:Smooth dependence on the input spike times} shows smooth dependence of later states and spike times on the spike arrival times, which form the endpoints of the intervals. The two remaining subsections, \cref{sec:Changing input spike order} and \cref{sec:Changing input and output spike order}, generalize the obtained results to neurons where the input spike times can change order with each other and with output spike times.

\subsection{Proof overview}\label{sec:proof overview}
For the proof it is helpful to transform $V(t)$ smoothly and bijectively to a phase variable $\phi(t)$ on a circle, i.e.~we transform the QIF to a $\theta$-neuron~\cite{EK86,latham2000intrinsic,I07,gerstner2014neuronal}. The momentary impact of the input current on the phase is then phase-dependent. The point of spike generation is in the $\phi$-dynamics not special anymore, except for the fact that the impact of the input current becomes zero there. This means that the threshold crossing itself happens purely due to the intrinsic neuron dynamics and always with the same finite rate of change $\dot{\phi}$. 

We start by considering the case where there are no input spikes and the initial conditions are varied. This entails the case of having a single input spike with varying weight (main text Fig.~1 left column). Assuming the neuron does spike at least once, the implicit function theorem~\cite{rudin1976principles} (thm.~9.28) together with the finite rate of change of $\phi$ at threshold crossing then implies that also the output spike times vary smoothly. The important difference to the LIF is here the always positive rate of change of $\phi$ at threshold crossing, which hinders the (dis-)appearance of spikes in the middle of a trial and that the gradient tends to infinity upon changing $w$.

Next, we consider the case of multiple input spikes with varying weights and times. If no spikes (two input or an input and an output spike), change order, the neuron's state prior to a given output spike but after the previous spike depends smoothly on the input parameters due to the smooth neuron dynamics between spikes. The considered output spike time then depends smoothly on this state because of the argument made above. More care has to be taken if two spikes change order. However, the dependence of output spike times turns out to be nevertheless smooth. For two interchanging input spikes this is ultimately because the order in which simultaneous inputs are processed does not matter (as they simply add to the current $I$). If an input and an output spike change order (main text Fig.~1 right column), it is because the impact of the input current on $\phi$ vanishes at the time of spike generation, as mentioned above. This is an important difference to the LIF and hinders the (dis-)appearance of spikes in the middle of a trial.

\subsection{Smooth dependence of the spike times on previous states}\label{sec:Smooth dependence on the initial conditions}

In this subsection we consider a scenario similar to main text Fig.~1 left column, i.e.~a QIF, \cref{eq:QIF general}, with an exponentially decaying input current,
\begin{align}
\label{eq:I dot single}
    \taus \dot{I}(t)&=-I(t),
\end{align}
for $t\geq 0$. The input may just have arrived at $t=0$. The parameters are the initial states, $V(0)=V_0$ and $I(0)=w$, which shall be both finite. 
We show that the states and output spike times depend smoothly on the parameters and that the output spikes appear for increasing input strength $w$ at infinite time or at the end of the trial, $T$, if it is earlier.

For this, we transform $V$ to an angle variable $\phi$ using \cref{eq:Phi}. At the point of threshold crossing, $\phi=1$, which is the same state as $\phi=0$. $\phi$'s temporal derivative at this point equals $1$, independent of $I$ and thus $w$.
We further restrict $w$ to some compact interval $[w_{\min},w_{\max} ]$ with $w_{\min}\leq 0 \leq w_{\max}$. The dynamics of $\phi$ and $I$ are for $t>0$ given by the smooth system of differential equations \cref{eq:phi dot,eq:I dot single}, which is defined on the compact set $S^1\times [w_{\min},w_{\max} ]$. The dynamics do not leave this set.
The solutions $\left(\begin{smallmatrix}\phi(t)\\ I(t)\end{smallmatrix}\right)$ thus exist for all times and depend smoothly on $t$ and the initial conditions $\phi(0)=\phi_0=\Phi(V_0)$ and $I(0)=w$, \cite{hirsch1974differential} (Secs.~8.5, 15.2), \cite{jetschke2009mathematik} (Sec.~1.2.3), \cite{arnold1992ordinary} (Sec.~5.35).

\begin{figure}
    \centering
    \includegraphics[width=1.\columnwidth]{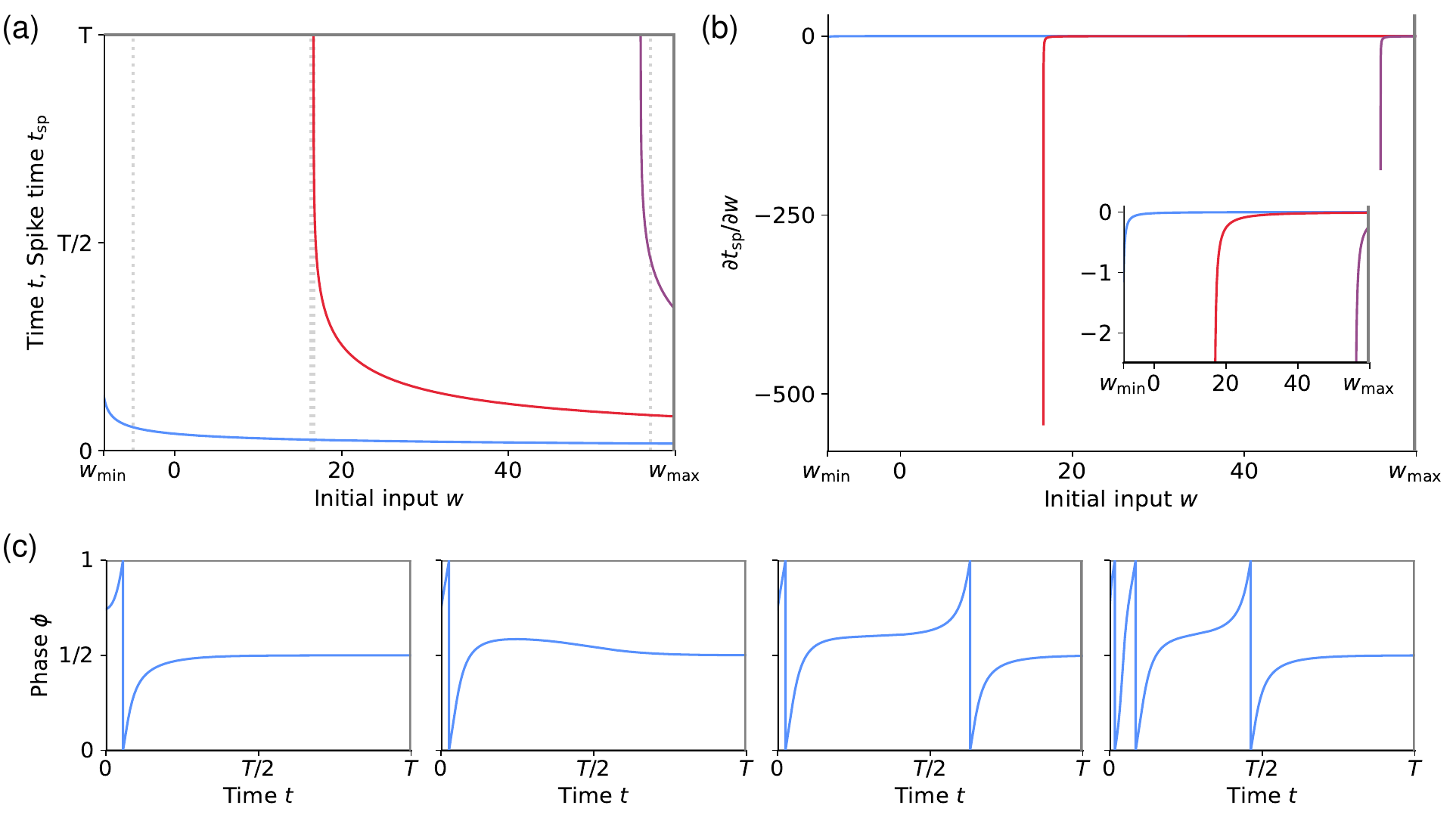}
    \caption{\label{fig:tsp(w)} Spike times of a QIF with a single exponentially decaying input arriving at $t=0$. (a) The output spike times $\tspike$ of the QIF form continuous curves without kinks in $(w,t)$-space (blue, red, purple: first, second, third output spike time), which start at $T$ or at $w_\text{min}$ and end at $w_\max$  ($T=10$, i.e.~ten times the membrane time constant, $w_\text{min}=-8.5, w_\max=60$). They are the graphs of smooth functions $\tspike(w)$. (b) Derivative of the output spike times with respect to $w$ (blue, red, purple: derivative of first, second, third output spike time). $\frac{\partial \tspike}{\partial w}$ is continuous. All derivative graphs start at finite values of $\partial \tspike/\partial w$, since the trial duration $T$ is finite. Starting points with $w>w_\text{min}$ correspond to points where $\tspike(w)$ starts to fall below $T$. Near these points, the derivatives assume large negative values. (c) Example traces $\phi(t)$ for different values of $w$ (from left to right: $w=-5, 16.3, 16.7, 57$, highlighted by light gray vertical dotted lines in (a)) show first one and then a second and third spike. Spikes appear at the end of the trial and then shift to earlier times with increasing $w$.}
\end{figure}

Interpreting $\phi$ for fixed $\phi_0$ as a function on $(w,t)$-space, we observe that the points $(w,t)$ mapped to $1$ specify the spike times $t=\tspike$ of the neuron for the input strengths $w$, \cref{fig:tsp(w)}a. Since $\{1\}$ is a closed set and $\phi$ continuous, the preimage of $\{1\}$, i.e.~the set of points $(w,t)$ mapped by $\phi$ to ${1}$, is closed as well. From the previous paragraph, we know that $\phi$ is even smooth in $t$ and $w$ and that the partial derivative with respect to $t$ is at spike times invertible, since $\partial \phi/\partial t|_{(w,\tspike)}=1\neq 0$. The implicit function theorem~\cite{rudin1976principles} (thm.~9.28) thus ensures that the set of spike times in $(w,t)$-space looks locally, around each of its points, like the graph of a smooth function $\tspike(w)$. The set thus consists of possibly multiple curves (for multiple spikes) in $(w,t)$-space, which are continuous, without ``kinks'' and with finite slope $d\tspike/dw$, except where $\tspike$ tends to infinity, \cref{fig:tsp(w)}a. The appearance of a spike corresponds to the start of such a curve. This start cannot lie in the interior of $(w,t)$-space, because the closeness implies that the starting point is part of the curve such that the implicit function theorem would guarantee continuation of the curve to both sides. The curves must thus extend to the borders of $(w,t)$-space. Specifically for growing $w$ they start at $t=\infty$ or $T$ or they start at $w=w_{\min}$, if $\phi_0$ is so large that the spike is generated already for this input weight. They end at $w=w_{\max}$, because the spike times decrease monotonically with $w$, as $\dot{\phi}$ increases with increasing input.
Spikes can for increasing $w$ therefore only appear at $t=\infty$ or $t=T$.  We note that the above argument also excludes merger of spike times, which would correspond to merger of curves.  Further, an alike argument shows that $\tspike$ depends smoothly on $\phi_0,w$, \cref{fig:tsp(w)}b. The closed sets mapped by $\phi$ to ${1}$ are then planes in $\phi_0,w,t$-space. The above arguments do not apply to LIFs, since the temporal derivative of the voltage can become zero at spike times, see main text Fig.~1 left column.

\subsection{Smooth dependence on input weights}\label{sec:Smooth dependence on the input weights}
The previous subsection showed that the membrane potential dynamics and the spike times of a QIF with an exponentially decaying input depend smoothly on the initial conditions $\phi_0$ and $I(0)$. We now turn to the case of multiple inputs and show smooth dependence of the output spike times $\tspike$ on the synaptic input weights. If multiple spikes arrive, the input current \cref{eq:I dot single} changes in a jump-like manner by $w_i$ at each arrival time $t_i$ of a spike from neuron $i$,
\begin{align}\label{eq:I dot}
    \taus \dot{I}(t)=-I(t)+ \taus \sum_i w_i \sum_{t_i}\delta(t-t_i).
\end{align}
Note that for simplicity we use $t_i$ for a single input spike, for all input spikes from neuron $i$ and for input spikes in general.
The jump-like change in $I$ renders the value of $I$ directly at $t_i$ undefined, such that we need to separately consider the limits from below and above, $I(t_i^-)$ and $I(t_i^+)$. Further, it leads to finite size jumps in the temporal derivative of $\phi$, but the value of $\phi$ itself still changes continuously. The previous subsection tells us that within the interval given by two subsequent spike times, $t_i$ and $t_j$, the state and possible spike times depend smoothly on the state in its beginning, $\left(\begin{smallmatrix}\phi(t_i)\\I(t_i^+)\end{smallmatrix}\right)$. This state results smoothly from the state at the end of the previous interval and the input weight, $\left(\begin{smallmatrix}\phi(t_i)\\I(t_i^+)\end{smallmatrix}\right)=\left(\begin{smallmatrix}\phi(t_i)\\I(t_i^-)+w_i\end{smallmatrix}\right)$. $\left(\begin{smallmatrix}\phi(t_i)\\I(t_i^-)\end{smallmatrix}\right)$, in turn, depends smoothly on the state at the beginning of the previous interval and so on.  Thus, the state at any time $t$ depends smoothly on the initial conditions at the very beginning and on the individual input weights. This implies that the partial derivatives of $\left(\begin{smallmatrix}\phi(t_i)\\I(t_i^+)\end{smallmatrix}\right)$ with respect to each $w_i$ are continuous. This holds irrespective of whether and when output spikes are generated, since the states where this happens, i.e.~where $\phi(\tspike)=1$ holds, are not special for the neuron dynamics in $\phi,I$-space.
A function is continuously differentiable in all its variables exactly if all partial derivatives exist and are continuous~\cite{rudin1976principles} (thm.~9.21). This implies that because $\left(\begin{smallmatrix}\phi\\I\end{smallmatrix}\right)$ is a smooth function of each single $w_i$, it is a smooth function of all $w_i$. For an output spike time $\tspike\neq t_j$ for all $j$, \cref{sec:Smooth dependence on the initial conditions} shows that $\tspike$ depends smoothly on closely nearby, previous states with no spike arrivals in between. The output spike time therefore also depends smoothly on all $w_i$. (If an input spike time agrees with an output spike time, $\tspike=t_j$, the state is discontinuous in time as there is a jump in the current. We will see in \cref{sec:Changing input and output spike order} that this does not cause problems, because the impact of inputs on $\phi$ vanishes at spike times.)

If a neuron receives at multiple times $t_i$ input from the same input neuron $i$, the additive changes in $I$ are the same, $w_i$, at these times. We have shown smooth dependence of $\tspike\neq t_j$ for all $j$ on the input weights of all input times, as if they were distinct variables. If some of these distinct variables have the same values and change in the same manner, $\tspike$ still changes smoothly, which ensures smooth dependence of the output on the actual $w_i$.

\subsection{Smooth dependence on input spike times}\label{sec:Smooth dependence on the input spike times}
In our gradient descent scheme, also the input spike times to a neuron may change, for example because they are the output spike times of other neurons in the network. In the following we show that the output spike times of a neuron depend smoothly on the input spike times, if the order of (input and output) spike times stays the same. Since $\left(\begin{smallmatrix}\phi\\I\end{smallmatrix}\right)$ depends smoothly on time between interval borders, $\left(\begin{smallmatrix}\phi(t_i)\\I(t_i^-)\end{smallmatrix}\right)$ depends smoothly on $t_i$. The same holds for $\left(\begin{smallmatrix}\phi(t_i)\\I(t_i^+)\end{smallmatrix}\right)$, since it differs from $\left(\begin{smallmatrix}\phi(t_i)\\I(t_i^-)\end{smallmatrix}\right)$ only by a constant shift by $w_i$ in $I$. Also the following states $\left(\begin{smallmatrix}\phi(t)\\I(t)\end{smallmatrix}\right)$, $t > t_i$, and thus (cf.~\cref{sec:Smooth dependence on the initial conditions}) the following output spike times $\tspike\neq t_j$ then depend smoothly on $t_i$. For a preceding state (at a time $t < t_i$) and for preceding output spike times the smoothness property is trivially satisfied, since there is no dependence on $t_i$. (If $t$ happens to agree with $t_i$, the state does not depend smoothly on $t_i$, because of the jump-like change in $I$.) Thus, as long as the output spike times satisfy $\tspike \neq t_i$ and $\tspike \neq t_j>t_i$, they depend smoothly on $t_i$, since there will always be states that depend smoothly on $t_i$ so closely before $\tspike$ that we can apply \cref{sec:Smooth dependence on the input weights}. (We note that since the times and states where an output spike is generated are not special for the neuron dynamics in $\phi,I$-space, the agreement of other spike times $t_j$ with other output spike times again does not change this.) Using also the results of the previous subsection, we conclude that as long as the spike order is conserved ($t_i\neq t_j$, $t_i\neq \tspike$, $t_j\neq \tspike$), states $\left(\begin{smallmatrix}\phi(t)\\I(t)\end{smallmatrix}\right)$, $t\neq t_i$, and thus also the output spike times are a smooth function of each single $w_i$ and $t_i$ and thus of all of them.

\subsection{Changing input spike order}\label{sec:Changing input spike order}

\begin{figure}
    \centering
    \includegraphics[width=1.\columnwidth]{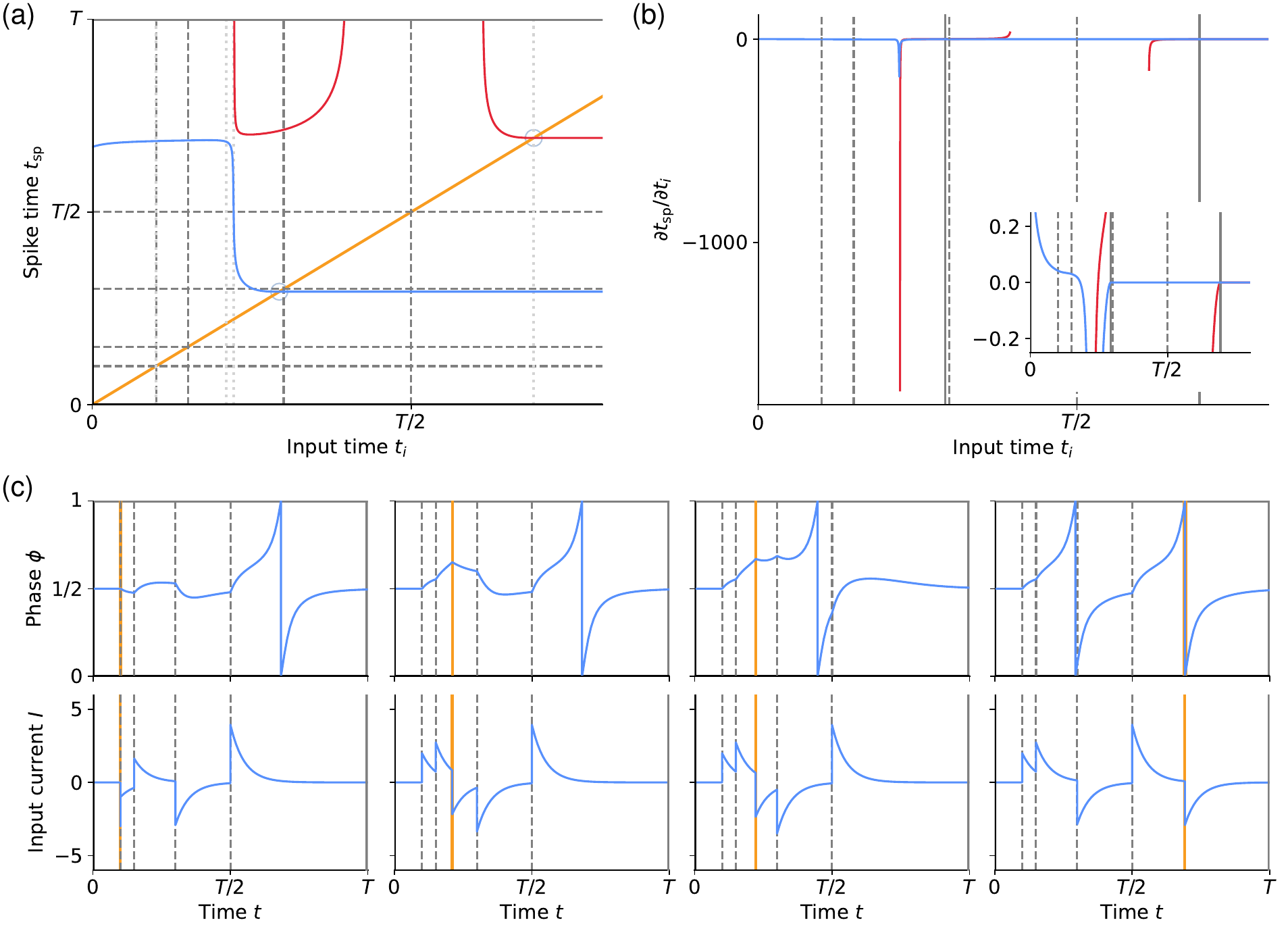}
    \caption{\label{fig:tsp(ti)} Change of output spike times when an input time changes. (a) The output spike times $\tspike$ (blue, red: first, second output spike time) are smooth functions of the input spike time $t_i$. There are no jumps or kinks in the graphs, also if $t_i$ crosses other input spike times $t_j$ (gray dashed vertical lines: $t_i=t_j$) or if it crosses output spike times (orange diagonal: $\tspike=t_i$, gray circles: crossing points of actual output spike times with $t_i$) or if the output spike times cross other input spike times $t_j$ (gray dashed horizontal lines: $\tspike=t_j$, partially crossed by blue curve). (b) The derivative $\partial \tspike/\partial t_i$ confirms the smoothness of the function $\tspike(t_i)$: It is continuous also at points where $t_i$ crosses other $t_j$ (gray dashed vertical lines) or where it agrees with actual output spike times (gray vertical lines). Inset: magnification of the range where derivatives are small, highlighting in particular the zero derivative when $t_i$ is larger than $\tspike$. The curves start and end at $w$ values where $\tspike(w)$ enters or exits the trial. (c) Example traces of $\phi(t)$ (upper panels) and $I(t)$ (lower panels) at different salient $t_i$ values (highlighted by light gray dotted lines in (a); gray dashed vertical lines in (c): $t_j$, orange vertical line: $t_i$): at the crossing of $t_i$ and a $t_j$ (trace one: $t_i=1$), closely before and after a fast change in the first spike time preceding an entering of the second spike time (traces two and three: $t_i=2.1, 2.22$) and close to the crossing of the second output spike time and $t_i$ (last trace: $t_i=6.92$).}
\end{figure}

This subsection investigates whether we have smooth dependence of the output spike times on the input spike times when the order of the input spike times changes.
Since the times of input and output spikes (henceforth, in short: events) form one-dimensional curves as a function of the training progress, interchanges of event order will generically happen, cf.~\cref{fig:tsp(ti),fig:tsp(wi)}. At a single point in the process, however, generically only two events cross. Therefore it suffices to only consider such cases here and in \cref{sec:Changing input and output spike order}. Specifically the current subsection shows that the state at a test time $t_2$, which is so close after a pair of spikes $t_i$ and $t_j$ that there is no further input time between them, depends smoothly on $t_i$ even if $t_i$ just changes order with $t_j$, i.e.~at $t_i=t_j$. Together with the results of \cref{sec:Smooth dependence on the input weights} (smooth dependence of subsequent on current states), \cref{sec:Smooth dependence on the initial conditions} (smooth dependence of output spike times on sufficiently closely preceding states), and \cref{sec:Smooth dependence on the input spike times} (smooth dependence of the output spike times on $t_i$ for $t_i\neq t_j$), this shows that the output spike times $\tspike$ depend smoothly on a single $t_i$, if they do not change order with it, i.e.~for all $t_i\neq \tspike$. From \cite{rudin1976principles} (thm.~9.2), we again conclude that the output spike times depend smoothly on all $t_i$, as long as $t_i\neq \tspike$. 
 
For our considerations, it is convenient to introduce some further notions and abbreviations. First, we will use the flow~\cite{jetschke2009mathematik,arnold1992ordinary,hirsch1974differential} generated by the free differential equations \cref{eq:phi dot,eq:I dot single}. This maps the state at $t_a$ to the state at $t_b$, if there are no input spikes arriving in between. We denote the flow by $T_{t_b-t_a}\left(\begin{smallmatrix}\phi(t_a)\\I(t_a^+)\end{smallmatrix}\right)$, such that
\begin{align}\label{eq:flow maps states}
\left(\begin{smallmatrix}\phi(t_b)\\I(t_b^-)\end{smallmatrix}\right)=T_{t_b-t_a}\left(\begin{smallmatrix}\phi(t_a)\\I(t_a^+)\end{smallmatrix}\right).
\end{align}
We know from \cref{sec:Smooth dependence on the initial conditions} that this flow is a smooth, vector-valued function of its time and state argument. While $\phi$ is a continuous function of time, $I$ is discontinuous at spike arrival times. Therefore, we regularly need to specify the left or right hand side time limits in the time argument of $I$ as indicated: if $t_a$ and $t_b$ are subsequent spike arrival times, $T_{t_b-t_a}$ maps the state directly after $t_a$ to the state directly before $t_b$. We further introduce the abbreviation $f$ for the right hand side of the system of differential equations \cref{eq:phi dot,eq:I dot single} to compactly write

\begin{align}\label{eq:f introduction}
    \left(\begin{smallmatrix}\dot{\phi}\\\dot{I}\end{smallmatrix}\right)=f\left(\begin{smallmatrix}\phi\vphantom{\dot{\phi}}\\I\vphantom{\dot{I}}\end{smallmatrix}\right).
\end{align}
As an immediate consequence of \cref{eq:flow maps states} the time derivative of the flow is
\begin{align}\label{eq:flow time derivative}
    \dot{T}_{t_b-t_a}\left(\begin{smallmatrix}\phi(t_a)\\I(t_a^+)\end{smallmatrix}\right)=f\left(\begin{smallmatrix}\phi(t_b)\\I(t_b^-)\end{smallmatrix}\right).
\end{align}
We will further use the derivative of the flow with respect to its state argument, the differential
\begin{align}\label{eq:flow differential}
    D{T}_{t_b-t_a}\left(\begin{smallmatrix}\phi(t_a)\\I(t_a^+)\end{smallmatrix}\right)=
    \left(\begin{matrix}
        &\frac{\partial\phi(t_b)}{\partial\phi(t_a)}
        &\frac{\partial \phi(t_b)}{\partial I(t_a^+)}\\&\frac{\partial I(t_b^-)}{\partial\phi(t_a)}&
        \frac{\partial I(t_b^-)}{\partial I(t_a^+)}
    \end{matrix}\right).
\end{align}

Our aim is to show the smooth dependence of $\left(\begin{smallmatrix}\phi(t_2)\\I(t_2)\end{smallmatrix}\right)$ on $t_i$ at $t_i=t_j$. For this we first show the continuity of $\left(\begin{smallmatrix}\phi(t_2)\\I(t_2)\end{smallmatrix}\right)$ and then the continuity of $\frac{\partial}{\partial t_i}\left(\begin{smallmatrix}\phi(t_2)\\I(t_2)\end{smallmatrix}\right)$ as a function of $t_i$, at $t_i=t_j$. We start by considering the dynamics at $t_1<\min(t_j,t_i)$, which shall be so close to $t_i,t_j$ that there are no further input times between them, similar to $t_2>\max(t_j,t_i)$. Three cases need to be distinguished: $t_i<t_j$, $t_i=t_j$ and $t_i>t_j$.
In the first case, the state at $t_2$ may be written as
\begin{align}\label{eq:state ti smaller tj}
    \left.\left(\begin{smallmatrix}
    \phi(t_2)\\
    I(t_2)
    \end{smallmatrix}\right)\right|_{t_i<t_j}
    =
    T_{t_2-t_j}\bigg\{
    T_{t_j-t_i}\bigg(
    T_{t_i-t_1}\bigg[\left(\begin{smallmatrix}
    \phi(t_1)\\
    I(t_1)
    \end{smallmatrix}\right)
    \bigg]
    +\left(\begin{smallmatrix}
    0\vphantom{\phi(t_1)}\\
    w_i\vphantom{I(t_1)}
    \end{smallmatrix}\right)
    \bigg)
    +\left(\begin{smallmatrix}
    0\vphantom{\phi(t_1)}\\
    w_j\vphantom{I(t_1)}
    \end{smallmatrix}\right)
    \bigg\},
\end{align}
in the second as
\begin{align}\label{eq:state ti equal tj}
    \left.\left(\begin{smallmatrix}
    \phi(t_2)\\
    I(t_2)
    \end{smallmatrix}\right)\right|_{t_i=t_j}
    =
    T_{t_2-t_j}\bigg\{
    T_{t_j-t_1}\bigg[\left(\begin{smallmatrix}
    \phi(t_1)\\
    I(t_1)
    \end{smallmatrix}\right)
    \bigg]
    +\left(\begin{smallmatrix}
    0\vphantom{\phi(t_1)}\\
    w_i+w_j\vphantom{I(t_1)}
    \end{smallmatrix}\right)
    \bigg\},
\end{align}
and in the third as
\begin{align}\label{eq:state ti larger tj}
    \left.\left(\begin{smallmatrix}
    \phi(t_2)\\
    I(t_2)
    \end{smallmatrix}\right)\right|_{t_i>t_j}
    =
    T_{t_2-t_i}\bigg\{
    T_{t_i-t_j}\bigg(
    T_{t_j-t_1}\bigg[\left(\begin{smallmatrix}
    \phi(t_1)\\
    I(t_1)
    \end{smallmatrix}\right)
    \bigg]
    +\left(\begin{smallmatrix}
    0\vphantom{\phi(t_1)}\\
    w_j\vphantom{I(t_1)}
    \end{smallmatrix}\right)
    \bigg)
    +\left(\begin{smallmatrix}
    0\vphantom{\phi(t_1)}\\
    w_i\vphantom{I(t_1)}
    \end{smallmatrix}\right)
    \bigg\}.
\end{align}
Here and in the following we employ also edged and curly brackets around function arguments to better distinguish them. To see the continuity of 
$\left(\begin{smallmatrix}\phi(t_2)\\I(t_2)\end{smallmatrix}\right)$ as a function of $t_i$ at $t_i=t_j$, we show that $\left(\begin{smallmatrix}\phi(t_2)\\I(t_2)\end{smallmatrix}\right)$ 
converge to the same state, $\left.\left(\begin{smallmatrix}
\phi(t_2)\\
I(t_2)
\end{smallmatrix}\right)\right|_{t_i=t_j}$, when $t_i$ approaches $t_j$ from below or above,
\begin{align}
    \lim_{t_i\nearrow t_j}
    \left.
    \left(\begin{smallmatrix}
    \phi(t_2)\\
    I(t_2)
    \end{smallmatrix}\right)
    \right|_{t_i<t_j}
    &=
    T_{t_2-t_j}\bigg\{
    T_{0}\bigg(
    T_{t_j-t_1}\bigg[\left(\begin{smallmatrix}
    \phi(t_1)\\
    I(t_1)
    \end{smallmatrix}\right)
    \bigg]
    +\left(\begin{smallmatrix}
    0\vphantom{\phi(t_1)}\\
    w_i\vphantom{I(t_1)}
    \end{smallmatrix}\right)
    \bigg)
    +\left(\begin{smallmatrix}
    0\vphantom{\phi(t_1)}\\
    w_j\vphantom{I(t_1)}
    \end{smallmatrix}\right)
    \bigg\}\\
    &=
    T_{t_2-t_j}\bigg\{
    T_{t_j-t_1}\bigg[\left(\begin{smallmatrix}
    \phi(t_1)\\
    I(t_1)
    \end{smallmatrix}\right)
    \bigg]
    +\left(\begin{smallmatrix}
    0\vphantom{\phi(t_1)}\\
    w_i+w_j\vphantom{I(t_1)}
    \end{smallmatrix}\right)
    \bigg\}\\
    &=\left.\left(\begin{smallmatrix}\phi(t_2)\\I(t_2)\end{smallmatrix}\right)\right|_{t_i=t_j}\\
    &=
    T_{t_2-t_j}\bigg\{
    T_{0}\bigg(
    T_{t_j-t_1}\bigg[\left(\begin{smallmatrix}
    \phi(t_1)\\
    I(t_1)
    \end{smallmatrix}\right)
    \bigg]
    +\left(\begin{smallmatrix}
    0\vphantom{\phi(t_1)}\\
    w_j\vphantom{I(t_1)}
    \end{smallmatrix}\right)
    \bigg)
    +\left(\begin{smallmatrix}
    0\vphantom{\phi(t_1)}\\
    w_i\vphantom{I(t_1)}
    \end{smallmatrix}\right)
    \bigg\}\\
    &=\lim_{t_i\searrow t_j}\left.\left(\begin{smallmatrix}
    \phi(t_2)\\
    I(t_2)
    \end{smallmatrix}\right)\right|_{t_i>t_j}.\label{eq:cont state ti=tj}
\end{align}
The first, third and fifth line uses \cref{eq:state ti smaller tj}, \cref{eq:state ti equal tj}  and \cref{eq:state ti larger tj}, respectively. The second and fourth lines use that the addition of weights commutes and that $T_0$ is the identity.
The continuity at $t_i=t_j$ is thus a consequence of the fact that the addition of inputs to the current is commutative.

We will proceed similarly to see the continuity of the partial derivative $\frac{\partial}{\partial t_i}\left(\begin{smallmatrix}\phi(t_2)\\I(t_2)\end{smallmatrix}\right)$. For this, we first compute the partial derivative for $t_i<t_j$ employing \cref{eq:state ti smaller tj}, the chain rule and \cref{eq:flow time derivative},
\begin{align}
    \frac{\partial}{\partial t_i}\left.\left(\begin{smallmatrix}\phi(t_2)\\I(t_2)\end{smallmatrix}\right)\right|_{t_i<t_j}
    &=DT_{t_2-t_j}\left(\begin{smallmatrix}\phi(t_j)\\I(t_j^+)\end{smallmatrix}\right)
    \cdot\bigg\{
    -\dot{T}_{t_j-t_i}\left(\begin{smallmatrix}\phi(t_i)\\I(t_i^+)\end{smallmatrix}\right)
    +DT_{t_j-t_i}\left(\begin{smallmatrix}\phi(t_i)\\I(t_i^+)\end{smallmatrix}\right)
    \cdot\bigg(\dot{T}_{t_i-t_1}\left(\begin{smallmatrix}\phi(t_1)\\I(t_1)\end{smallmatrix}\right)\bigg)
    \bigg\}\\
    &=DT_{t_2-t_j}\left(\begin{smallmatrix}\phi(t_j)\\I(t_j^+)\end{smallmatrix}\right)
    \cdot\bigg\{
    -f\left(\begin{smallmatrix}\phi(t_j)\\I(t_j^-)\end{smallmatrix}\right)
    +DT_{t_j-t_i}\left(\begin{smallmatrix}\phi(t_i)\\I(t_i^+)\end{smallmatrix}\right)\cdot f\left(\begin{smallmatrix}\phi(t_i)\\I(t_i^-)\end{smallmatrix}\right)
    \bigg\}.\label{eq:partial deriv ti smaller tj}
\end{align}
For $t_i>t_j$, we obtain from \cref{eq:state ti larger tj}
\begin{align}
    \left.\frac{\partial}{\partial t_i}\left(\begin{smallmatrix}\phi(t_2)\\I(t_2)\end{smallmatrix}\right)\right|_{t_i>t_j}
    &=-\dot{T}_{t_2-t_i}\left(\begin{smallmatrix}\phi(t_i)\\I(t_i^+)\end{smallmatrix}\right)
    +DT_{t_2-t_i}\left(\begin{smallmatrix}\phi(t_i)\\I(t_i^+)\end{smallmatrix}\right)
    \dot{T}_{t_i-t_j}\left(\begin{smallmatrix}\phi(t_j)\\I(t_j^+)\end{smallmatrix}\right)\\
    &=-DT_{t_2-t_i}\left(\begin{smallmatrix}\phi(t_i)\\I(t_i^+)\end{smallmatrix}\right)
    f\left(\begin{smallmatrix}\phi(t_i)\\I(t_i^+)\end{smallmatrix}\right)
    +DT_{t_2-t_i}\left(\begin{smallmatrix}\phi(t_i)\\I(t_i^+)\end{smallmatrix}\right)
    \dot{T}_{t_i-t_j}\left(\begin{smallmatrix}\phi(t_j)\\I(t_j^+)\end{smallmatrix}\right)
    \\
    &=DT_{t_2-t_i}\left(\begin{smallmatrix}\phi(t_i)\\I(t_i^+)\end{smallmatrix}\right)
    \cdot\bigg(
    -f\left(\begin{smallmatrix}\phi(t_i)\\I(t_i^+)\end{smallmatrix}\right)
    +
    f\left(\begin{smallmatrix}\phi(t_i)\\I(t_i^-)\end{smallmatrix}\right)
    \bigg).\label{eq:partial deriv ti larger tj}
\end{align}
The second line uses the general relation
\begin{align}
    \dot{T}_t(x)&=\left.\frac{d}{dr}T_{t+r}(x)\right|_{r=0}=\frac{d}{dr}\left.T_{t}\bigg(T_r(x)\bigg)\right|_{r=0}\\
    &=DT_{t}(x)\cdot \dot{T}_r(x)\vphantom{\big(}\bigg|_{r=0}=DT_{t}(x)\cdot f(x).
\end{align}
It reflects that we obtain the same state change if we (i) evolve the system about an infinitesimal interval $dt$ past $t$ (state change $\dot{T}_t(x)dt$) or if we (ii) evolve the initial state about $dt$ (state change $f(x)dt$) and then evolve the change about $t$ (via $DT_{t}(x)$, linear approximation suffices).
We now compare the values of $\frac{\partial}{\partial t_i}\left(\begin{smallmatrix}\phi(t_2)\\I(t_2)\end{smallmatrix}\right)$ when $t_i$ approaches $t_j$ from below or above. \cref{eq:partial deriv ti smaller tj} yields
\begin{align}
    \lim_{t_i\nearrow t_j}\frac{\partial}{\partial t_i}\left.\left(\begin{smallmatrix}\phi(t_2)\\I(t_2)\end{smallmatrix}\right)\right|_{t_i<t_j}
    &=
    DT_{t_2-t_j}\left(\begin{smallmatrix}\phi(t_j)\\I(t_j^-)+w_i+w_j\end{smallmatrix}\right)
    \cdot\bigg\{
    -f\left(\begin{smallmatrix}\phi(t_j)\\I(t_j^-)+w_i\end{smallmatrix}\right)
    +DT_{0}\left(\begin{smallmatrix}\phi(t_j)\\I(t_j^-)+w_i\end{smallmatrix}\right)\cdot f\left(\begin{smallmatrix}\phi(t_j)\\I(t_j^-)\end{smallmatrix}\right)
    \bigg\}\\
    &=DT_{t_2-t_j}\left(\begin{smallmatrix}\phi(t_j)\\I(t_j^-)+w_i+w_j\end{smallmatrix}\right)
    \cdot\bigg\{
    -f\left(\begin{smallmatrix}\phi(t_j)\\I(t_j^-)+w_i\end{smallmatrix}\right)
    +f\left(\begin{smallmatrix}\phi(t_j)\\I(t_j^-)\end{smallmatrix}\right)
    \bigg\}\label{eq:limit partial deriv ti smaller tj},
\end{align}
where the limit in $I(t_j^-)$ is taken after the after the limit $t_i\nearrow t_j$, such that $I(t_j^-)$ is the current at $t_j$ without both inputs $w_i$ and $w_j$. From \cref{eq:partial deriv ti larger tj} we obtain
\begin{align}
    \lim_{t_i\searrow t_j}\left.\frac{\partial}{\partial t_i}\left(\begin{smallmatrix}\phi(t_2)\\I(t_2)\end{smallmatrix}\right)\right|_{t_i>t_j}
    =
    DT_{t_2-t_j}\left(\begin{smallmatrix}\phi(t_j)\\I(t_j^-)+w_i+w_j\end{smallmatrix}\right)
    \cdot\bigg(
    -f\left(\begin{smallmatrix}\phi(t_j)\\I(t_j^-)+w_i+w_j\end{smallmatrix}\right)
    +
    f\left(\begin{smallmatrix}\phi(t_j)\\I(t_j^-)+w_j\end{smallmatrix}\right)
    \bigg).\label{eq:limit partial deriv ti larger tj}
\end{align}
The right hand side of the system of differential equations \cref{eq:phi dot,eq:I dot single} is an affine map in $I$. It has the form 
\begin{align}\label{eq:f1 f2}
    f\left(\begin{smallmatrix}\phi\\ I\end{smallmatrix}\right)=f_1(\phi)+f_2(\phi)I,
\end{align}
with vector valued functions $f_1(\phi)=\left(\begin{smallmatrix}\cos(\pi\phi)\left(\cos(\pi\phi)+\tfrac{1}{\pi}\sin(\pi\phi)\right)\\
0
\end{smallmatrix}\right)$ and $f_2(\phi)=\left(\begin{smallmatrix}\tfrac{1}{\pi^2}\sin^2(\pi \phi)\\
-\tfrac{1}{\taus}
\end{smallmatrix}\right)$. The limits in \cref{eq:limit partial deriv ti smaller tj} and \cref{eq:limit partial deriv ti larger tj} thus agree,
\begin{align}\label{eq:ti left and right derivative}
    \lim_{t_i\nearrow t_j}\frac{\partial}{\partial t_i}\left(\begin{smallmatrix}\phi(t_2)\\I(t_2)\end{smallmatrix}\right)=
    DT_{t_2-t_j}\left(\begin{smallmatrix}\phi(t_j)\\I(t_j^-)+w_i+w_j\end{smallmatrix}\right)
    \cdot\bigg\{
    -f_2(\phi(t_j))w_i
    \bigg\}
    =\lim_{t_i\searrow t_j}\frac{\partial}{\partial t_i}\left(\begin{smallmatrix}\phi(t_2)\\I(t_2)\end{smallmatrix}\right).
\end{align}
Together with the continuity of $\left(\begin{smallmatrix}\phi(t_2)\\I(t_2)\end{smallmatrix}\right)$ as a function of $t_i$ around $t_j$ (\cref{eq:cont state ti=tj}, \cref{sec:Smooth dependence on the input spike times}), this implies that the partial derivative $\frac{\partial}{\partial t_i}\left(\begin{smallmatrix}\phi(t_2)\\I(t_2)\end{smallmatrix}\right)$ exactly at $t_i=t_j$ exists~\cite{heuser98lehrbuch} (p.~286, Ex.~5) as well: it equals the limits \cref{eq:ti left and right derivative},
\begin{align}\label{eq:ti derivative at tj}
    \left.\frac{\partial}{\partial t_i}\left(\begin{smallmatrix}\phi(t_2)\\I(t_2)\end{smallmatrix}\right)\right|_{t_i=t_j}=
    DT_{t_2-t_j}\left(\begin{smallmatrix}\phi(t_j)\\I(t_j^-)+w_i+w_j\end{smallmatrix}\right)
    \cdot\bigg\{
    -f_2(\phi(t_j))w_i
    \bigg\}.
\end{align}
Intuitively the employed theorem indicates that a continuous function that is not differentiable has some kink; it can be proven using the mean value theorem~\cite{rudin1976principles} (thm.~5.10).
We conclude that $\left(\begin{smallmatrix}\phi(t_2)\\I(t_2)\end{smallmatrix}\right)$ depends smoothly on $t_i$ also if it crosses other input spike times.

Since we can choose $t_2$ arbitrarily close to $t_j$, all states $\left(\begin{smallmatrix}\phi(t)\\I(t)\end{smallmatrix}\right)$, $t\neq t_i$ depend smoothly on each and thus on all $t_i$, even if the input spikes change order with each other. As a consequence, the spike times $\tspike$ are a smooth function of all $t_i\neq \tspike$, even if these change order with each other, see \cref{fig:tsp(ti)}. This is again because there are always states so closely before $\tspike$ that there are no further input spikes in between, because these states depend smoothly on the $t_i$ and because $\tspike$ depends smoothly on them (\cref{sec:Smooth dependence on the initial conditions}).

\subsection{Changing input and output spike order}\label{sec:Changing input and output spike order}

In this subsection, we address input and output spike times that change order. This can happen because the input spikes change such that they cross output spikes and/or because the output spikes change (for example due to changes in previous input weights). Considering such crossings is particularly important, since also in a QIF an inhibitory input usually leads to a downward jump in the voltage derivative and thus to a downward kink in the voltage (main text Fig.~1 right column). This kink, however, vanishes when $t_i$ and $\tspike$ cross, preventing disruptive spike (dis-)appearances like in the LIF.

We consider an output spike time $\tspike$ that tends to agree or agrees with an input spike time $t_i$. We first show that $\tspike$ does not (dis-)appear in the middle of the trial and changes continuously and even smoothly as a function of previous $t_j<t_i$ and their weights $w_j$. For $t_j>t_i$ the property is obvious since there is no dependence on subsequent inputs, which is a special case of smooth dependence. Thereafter we show that $\tspike$ does not (dis-)appear in the middle of the trial and changes continuously and even smoothly when $t_i$ changes. $\tspike$ cannot (dis-)appear and changes smoothly with the weight $w_i$ associated with $t_i=\tspike$, since changes in the input current $I$ that take place at an output spike time leave the momentary phase $\phi(\tspike)=1$ and thus $\tspike$ unaffected, \cref{eq:phi dot}.

We first investigate whether $\tspike=t_i$ may disruptively (dis-)appear and whether it changes continuously when previous $t_j<t_i$ change. For this we note that in contrast to \cref{sec:Smooth dependence on the input weights,sec:Smooth dependence on the input spike times}, we cannot simply use the implicit function theorem (via \cref{sec:Smooth dependence on the initial conditions}) to determine the properties of $\tspike$, because at input arrivals $I(t)$ changes discontinuously, such that also $\phi(t)$ is not continuously differentiable with respect to time. 
This change, however, affects $\phi(t)$ only after $t_i$ (for $t>t_i$). Therefore, if $t_j$ tends to a limiting value $t_{j,0}$ such that $\tspike$ tends to $t_i$ from below, $\tspike$ behaves like an output spike in a system without input at $t_i$. In particular, it depends smoothly on $t_j$ and assumes the limiting value, $\tspike=t_i$, if $t_j$ assumes the limiting value, $t_j=t_{j,0}$ (\cref{sec:Smooth dependence on the input spike times}). If $t_j$ tends to $t_{j,0}$ such that $\tspike$ tends to $t_i$ from above, \cref{sec:Smooth dependence on the initial conditions} tells that $\phi(t_i)\nearrow 1$ and $\dot{\phi}(t_i^+)\rightarrow 1$, \cref{eq:phi dot}. This implies that in the limit there is a threshold crossing at $t_i$, i.e.~$\tspike=t_i$ for $t_j=t_{j,0}$. Therefore, output spikes tending to $t_i$ cannot vanish directly before reaching this limit, but continuously assume it. May an output spike vanish after reaching $t_i$, i.e.~when $\tspike=t_i$? To answer this we first note that the states $\left(\begin{smallmatrix}\phi(t)\\I(t)\end{smallmatrix}\right)$ with $t$ smaller or larger but sufficiently close (such that there are no further spike arrivals in between) to $t_i$, $t\lesssim t_i$ or $t\gtrsim t_i$, depend smoothly on $t$ 
(\cref{sec:Smooth dependence on the initial conditions}). The same holds for the time derivative $\dot{\phi}(t)$, because it is a smooth function of $\phi(t)$ and $I(t)$, \cref{eq:phi dot}. Further we know from \cref{sec:Smooth dependence on the input weights,sec:Smooth dependence on the input spike times} that the states $\left(\begin{smallmatrix}\phi(t)\\I(t)\end{smallmatrix}\right)$ and thus $\dot{\phi}(t)$ with $t\approx t_i$
depend smoothly on previous $t_j$. We again denote by $t_{j,0}$ the value of $t_j$ at which $\tspike=t_i$. For $t_j=t_{j,0}$, $\phi(t_i)=1$ and $\dot{\phi}(t_i^\pm)=1$: the impact of the input at $t_i$ vanishes and there is no kink in the phase despite the discontinuity of $I$ at $t_i$. Due to the above mentioned smooth dependence on $t$ we have $\dot{\phi}(t)\approx 1>0$ for $t\approx t_i$. Due to the smooth dependence of $\dot{\phi}(t)$ on $t_j$, it is positive also for $t_j\approx t_{j,0}$ and $\phi(t)$ is then a strictly monotonously increasing function of $t$ for $t\approx t_i$. Therefore there is at most one threshold crossing. Furthermore, the values of $\phi(t)$ are close to their values for $t_j=t_{j,0}$ and $\phi(t)$ is continuous as a function of $t$.
This guarantees a threshold crossing near $t_i$. We conclude that if there is a threshold crossing at $t_i$ for $t_j=t_{j,0}$, also if $t_j$ is unequal but sufficiently close to $t_{j,0}$ exactly one threshold crossing takes place, at a value $\tspike$ near $t_i$. Because $t_j\to t_{j,0}$ implies $\phi(t_i)\to 1$ and $\dot{\phi}(t_i^\pm)\to 1$, we have $\tspike\to t_i$, as already observed above.
The spike time $\tspike$ therefore does not disappear at $t_i$ and changes continuously with $t_j$. We conclude that spikes cannot (dis-)appear at or in the direct vicinity of an input spike $t_i$ due to continuous changes in previous input spike times. Furthermore output spike times $\tspike$ depend continuously on previous input spike times $t_j$ also if $\tspike$ agrees with an input spike time, $\tspike=t_i$.
We can see analogously that the same holds for the weights $w_j$ associated with $t_j$.

To show the existence and continuity of the derivative $\partial \tspike/\partial t_j$ at $t_{j,0}$, we compute the derivatives $\partial \tspike/\partial t_j$ for $\tspike$ being close to but smaller or larger than $t_i$, $\tspike\lesssim t_i$ or $\tspike\gtrsim t_i$. We will observe that they tend to the same limit if $t_j$ tends to $t_{j,0}$ such that $\tspike$ tends to $t_i$ from below or above. This implies existence and continuity of $\partial \tspike/\partial t_j$ and the limit yields the value of this derivative at $t_{j,0}$~\cite{heuser98lehrbuch} (p.~286, Ex.~5), cf.~also \cref{sec:Changing input spike order}, \cref{eq:ti derivative at tj}. If $\tspike\lesssim t_i$ or $\tspike\gtrsim t_i$, $\partial \tspike/\partial t_j$ can be computed using $\phi(\tspike)-1=0$ and the implicit function theorem,
\begin{align}\label{eq:spike time derivative}
\frac{\partial \tspike}{\partial t_j}&=-\frac{1}{\dot{\phi}(\tspike)}\frac{\partial \phi(\tspike)}{\partial t_j}=-\frac{\partial \phi(\tspike)}{\partial t_j},
\end{align}
where we have employed that always $\dot{\phi}(\tspike)=1$.
If we choose again a reference time $t_1$ that is sufficiently close before $t_i$ and $\tspike$, we obtain for $\tspike\lesssim t_i$,
\begin{align}
    \phi(\tspike)&=\left[T_{\tspike-t_1}\left(\begin{smallmatrix}
    \phi(t_1)\\
    I(t_1)
    \end{smallmatrix}\right)\right]_\phi,\\
    \frac{\partial \phi(\tspike)}{\partial t_j}
    &=\left[DT_{\tspike-t_1}\left(\begin{smallmatrix}
    \phi(t_1)\\
    I(t_1)
    \end{smallmatrix}\right)
    \cdot
    \left(\begin{smallmatrix}
    \frac{\partial\phi(t_1)}{\partial t_j}\\
    \frac{\partial I(t_1)}{\partial t_j}\
    \end{smallmatrix}\right)
    \right]_\phi,\\
    \label{eq:partial phi tsp tj from below}
    \lim_{\tspike\nearrow t_i}\frac{\partial \phi(\tspike)}{\partial t_j}
    &=\left[DT_{t_i-t_1}\left(\begin{smallmatrix}
    \phi(t_1)\\
    I(t_1)
    \end{smallmatrix}\right)
    \cdot
    \left(\begin{smallmatrix}
    \frac{\partial\phi(t_1)}{\partial t_j}\\
    \frac{\partial I(t_1)}{\partial t_j}\
    \end{smallmatrix}\right)
    \right]_\phi.
\end{align}
$[.]_\phi$ means that we only take the first, $\phi$-component of the final vector-valued expression. The limit in the last line occurs through $t_j$ tending appropriately to $t_{j,0}$.
If $\tspike\gtrsim t_i$, we analogously have
\begin{align}
    \phi(\tspike)&=\left[T_{\tspike-t_i}\left(T_{t_i-t_1}\left(\begin{smallmatrix}
    \phi(t_1)\\
    I(t_1)
    \end{smallmatrix}\right)+\left(\begin{smallmatrix}
    0\vphantom{\phi(t_1)}\\
    w_i\vphantom{I(t_1)}
    \end{smallmatrix}\right)\right)\right]_\phi,\\
    \frac{\partial \phi(\tspike)}{\partial t_j}
    &=
    \left[DT_{\tspike-t_i}\left(\begin{smallmatrix}
    \phi(t_i)\\
    I(t_i^+)
    \end{smallmatrix}\right)\cdot DT_{t_i-t_1}\left(\begin{smallmatrix}
    \phi(t_1)\\
    I(t_1)
    \end{smallmatrix}\right)
    \cdot
    \left(\begin{smallmatrix}
    \frac{\partial\phi(t_1)}{\partial t_j}\\
    \frac{\partial I(t_1)}{\partial t_j}\
    \end{smallmatrix}\right)
    \right]_\phi\\
    \label{eq:partial phi tsp tj from above}
    \lim_{\tspike\searrow t_i}\frac{\partial \phi(\tspike)}{\partial t_j}
    &=
    \left[DT_{t_i-t_1}\left(\begin{smallmatrix}
    \phi(t_1)\\
    I(t_1)
    \end{smallmatrix}\right)
    \cdot
    \left(\begin{smallmatrix}
    \frac{\partial\phi(t_1)}{\partial t_j}\\
    \frac{\partial I(t_1)}{\partial t_j}\
    \end{smallmatrix}\right)
    \right]_\phi.
\end{align}
In the last line we used that $DT_{0}\left(\begin{smallmatrix}
\phi(t_i)\\
I(t_i^+)
\end{smallmatrix}\right)$ is the identity matrix. The agreement 
of the partial derivatives' limits \cref{eq:partial phi tsp tj from below} and \cref{eq:partial phi tsp tj from above} reflects the fact that the states directly before and after $t_i$ only differ by an addition of $w_i$, which moreover occurs to $I$, not to $\phi$, such that the derivatives of $\phi(t_i^-)$ and $\phi(t_i^+)$ with respect to $t_j$ are the same. The agreement shows the smooth dependence of $\tspike$ on $t_j$ at $t_j=t_{j,0}$, where $\tspike=t_i$.
An analogous consideration shows the existence and continuity of the derivative $\partial \tspike/\partial w_j$ at $w_{j,0}$.
We conclude that $\tspike$ depends smoothly on earlier weights and spike times, also if it agrees with an input spike time.

\begin{figure}
    \centering
    \includegraphics[width=1.\columnwidth]{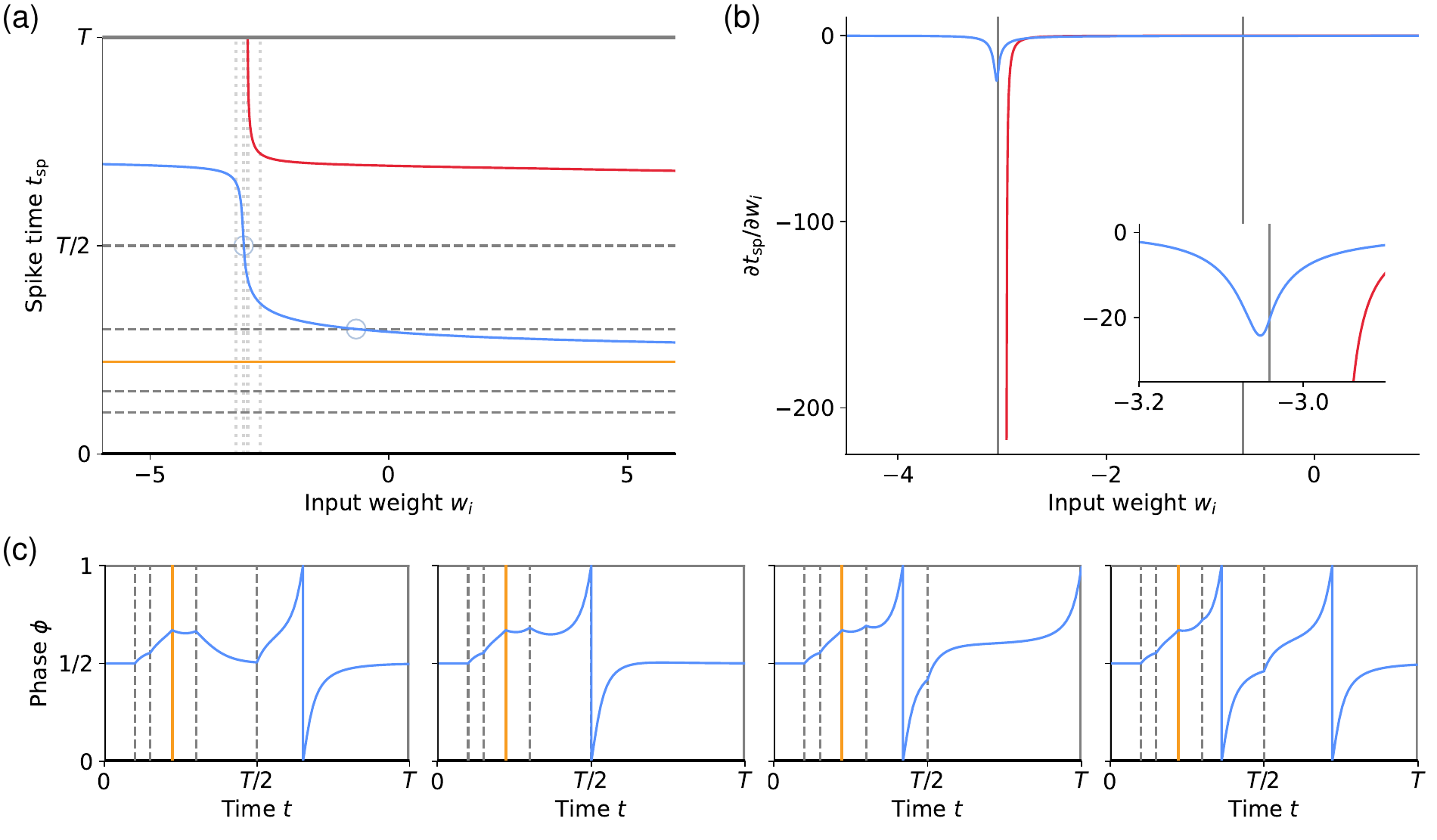}
    \caption{\label{fig:tsp(wi)} 
    Change of output spike times when the strength of one of multiple inputs changes. (a) The output spike times $\tspike$ (blue, red: first, second output spike time) are smooth functions of the input strength $w_i$ arriving at $t_i$ (orange horizontal line: $\tspike=t_i$). There are no jumps or kinks in the graphs, also when $\tspike$ crosses input spike times (gray dashed horizontal lines: $\tspike=t_j$, partially crossed by blue curve). (b) The derivative $\partial \tspike/\partial w_i$ confirms this smoothness. It is continuous also at values of $w_i$ where the output spike times cross input spike times (gray vertical lines; inset: magnification of the region around the crossing with smallest $w_i$). (c) Example traces of $\phi(t)$ at $w_i$ values around the fast change and the first crossing of the first $\tspike$ with a $t_j$ ($w_i=-3.2,-3.041,-2.957,-2.7$, highlighted by light gray dotted lines in (a); gray dashed vertical lines: $t_j$, orange vertical line: $t_i$).}
\end{figure}

We now study the only remaining case, the dependence of $\tspike$ on $t_i$ at $\tspike=t_i$. We first assume that  for $t_i$ tending to $t_{i,0}$ from below, there is a spike $\tspike$ tending to $t_{i,0}$. We ask if the spike will reach $t_{i,0}$ or whether it may disappear. This spike must occur after $t_i$, otherwise it cannot depend on $t_i$ and converge to $t_{i,0}>t_i$. This implies that $\phi(t_i)\lesssim 1$, because the threshold crossing with $\phi(\tspike)=1$ is a bit later than $t_i$, and $\phi(t_i)\nearrow 1$. Again because the input at $t_i$ affects the dynamics only after $t_i$, also in a modified system where this input is removed we have $\phi(t_i)\nearrow 1$ when $t_i\nearrow t_{i,0}$. In the modified system the phase dynamics are a smooth function of $t$ around $t_{i,0}$. Thus, in the limit $t_i=t_{i,0}$ we have $\phi(t_i)=1$ such that $t_{i,0}$ is a spike time of the modified system. If the input spike only arrives at $t_{i,0}$, the original and the modified systems' phases agree up to and including $t_{i,0}$. Therefore, also in the original system, we have $\phi(t_i)=1$ for $t_i=t_{i,0}$. This implies that if $\tspike$ tends to $t_{i,0}$ with $t_i\nearrow t_{i,0}$, $\tspike$ also reaches the limit, $\tspike=t_{i,0}$, for $t_i= t_{i,0}$.
Now we consider the case that $t_i$ tends to $t_{i,0}$ from above and $\tspike$ tends to $t_{i,0}$. Since $t_i>t_{i,0}$ cannot influence $\phi(t_{i,0})$, we must have $\phi(t_{i,0})=1$, so $\tspike=t_{i,0}$ for all the $t_i$ tending to $t_{i,0}$. Since also an input at $t_i$ does not change $\phi(t_{i})$, for $t_i=t_{i,0}$ we have $\phi(t_{i,0})=1$ as well.
$\tspike$ therefore cannot suddenly disappear in the vicinity of $t_{i,0}$ due to $t_i$ tending to and finally reaching $t_{i,0}$.
Can a spike suddenly (dis-)appear at $\tspike=t_{i,0}$ when $t_i=t_{i,0}$?
If $\tspike = t_i(=t_{i,0})$, the value of $\phi(t_{i,0})$ in the presence and in the absence of input at $t_i$ are equal, again because the input $w_i$ has no immediate impact on $\phi$. (Moreover, the impact of any input vanishes at $\phi(\tspike)=1$.) Therefore $t_{i,0}$ is a spike time if the input at $t_i$ is removed and for $t_i\geq t_{i,0}$.
In the latter case, $\tspike$ is constant as a function of $t_i$, in particular it does not vanish and depends smoothly on $t_i$. We thus consider $t_i\lesssim t_{i,0}$ in the following. In the absence of an input at $t_i$, the states sufficiently closely before $t_{i,0}$ are a smooth function of $t$. Further, since there is a threshold crossing at $t_{i,0}$, which implies a phase slope of $\dot{\phi}(t_{i,0})=1$, we have a phase that is slightly smaller than the threshold, $\phi(t)\lesssim 1$, for $t\lesssim t_{i,0}$. As a consequence, also in the system with input at $t_i$ we have for $t_i\nearrow t_{i,0}$ that $\phi(t_i)\nearrow 1$ and $\dot{\phi}(t_i^-)\to 1$, \cref{eq:phi dot}. Further, because the impact of an input goes to zero when approaching the threshold, we have $\dot{\phi}(t_i^+)\to 1$. The smoothness of the $\phi,I$-dynamics behind $t_i$ and the convergence to a nonzero $\dot{\phi}(t_i^+)$ implies that the $\phi$-dynamics will reach $1$ if the initial condition $\phi(t_i)$ is close enough to $1$. This shows that for $t_i$ sufficiently close to $t_{i,0}$ there will be a spike time $\tspike\approx t_{i,0}$. Therefore spikes cannot appear at $t_i=t_{i,0}$. Furthermore, the threshold crossing will be arbitrarily closely after $t_i$ for $\phi(t_i)$ tending to $1$. Therefore, the spike time $\tspike$ converges to $t_i$ and thus to $t_{i,0}$. We conclude that spikes cannot (dis-)appear at $\tspike=t_i=t_{i,0}$ and $\tspike$ is a continuous function of $t_i$ at $\tspike=t_i$. 

Also the partial derivative $\tfrac{\partial \tspike}{\partial t_i}$ is continuous at  $t_i=t_{i,0}$, where $\tspike=t_i$: We need to show that $\lim_{t_i\nearrow t_{i,0}} \tfrac{\partial \tspike}{\partial t_i}=0$, since $\lim_{t_i\searrow t_{i,0}} \tfrac{\partial \tspike}{\partial t_i}=0$ due to $\tspike$'s independence of $t_i$ for $t_i\geq t_{i,0}$ (where we have $\tspike=t_{i,0}$, see the previous paragraph). We again choose a reference time $t_1$ so close before $t_i\lesssim t_{i,0}$ that there are no further inputs in between and $t_i$ so close to $t_{i,0}$ that there is no spike arrival between $t_i$ and the spike time $\tspike\approx t_{i,0}$. Based on $\phi(\tspike)-1=0$ the implicit function theorem yields the derivative
\begin{align}\label{eq:spike time derivative ti}
    \frac{\partial \tspike}{\partial t_i}&=-\frac{1}{\dot{\phi}(\tspike)}\frac{\partial \phi(\tspike)}{\partial t_i}=-\frac{\partial \phi(\tspike)}{\partial t_i}\\
    &=-\frac{\partial}{\partial t_i}\left[T_{\tspike-t_i}\bigg(T_{t_i-t_1}
    \bigg[
    \left(\begin{smallmatrix}
    \phi(t_1)\\
    I(t_1)
    \end{smallmatrix}\right)
    \bigg]+\left(\begin{smallmatrix}
    0\vphantom{\phi(t_1)}\\
    w_i\vphantom{I(t_1)}
    \end{smallmatrix}\right)\bigg)
    \right]_\phi\\
    &=-
    \left[
    -\dot{T}_{\tspike-t_i}\left(\begin{smallmatrix}\phi(t_i)\\I(t_i^+)\end{smallmatrix}\right)
    +
    DT_{\tspike-t_i}\left(\begin{smallmatrix}\phi(t_i)\\I(t_i^+)\end{smallmatrix}\right)
    \cdot\bigg(\dot{T}_{t_i-t_1}\left(\begin{smallmatrix}\phi(t_1)\\I(t_1)\end{smallmatrix}\right)\bigg)
    \right]_\phi\\
    &=
    1
    -
    \left[DT_{\tspike-t_i}\left(\begin{smallmatrix}\phi(t_i)\\I(t_i^+)\end{smallmatrix}\right)
    \cdot f\left(\begin{smallmatrix}\phi(t_i)\\I(t_i^-)\end{smallmatrix}\right)
    \right]_\phi,
\end{align}
where we used in the last line that the $\phi$-component of $f$ is $1$ at a spike time, \cref{eq:phi dot}.
For $t_i\nearrow t_{i,0}$, also $\tspike$ tends to $t_{i,0}$, such that $DT_{\tspike-t_i}\left(\begin{smallmatrix}\phi(t_i)\\I(t_i^+)\end{smallmatrix}\right)$ becomes the identity matrix and the  $\phi$-component of $f\left(\begin{smallmatrix}\phi(t_i)\\I(t_i^-)\end{smallmatrix}\right)$ tends to its value at a spike time, $1$, since $t_i$ tends to a spike time of the dynamics. It follows that
\begin{align}
    \lim_{t_i\nearrow t_{i,0}}\frac{\partial \tspike}{\partial t_i}&=1-1=0.
\end{align}
This shows that $\tspike$ is a smooth function of $t_i$ also if $t_i$ crosses $\tspike$.

\newpage

\section{Pseudospike time smoothness and continuity for QIFs with extended coupling}

\subsection{First type of pseudospikes}\label{sec:QIF pseudo1 proof}

In this section, we prove that the pseudospike times for QIFs with extended coupling of the first type (\cref{sec:QIF pseudo1}), including their transitions to ordinary spike times, are continuous and mostly smooth in the network parameters (weights and input spike times). This is condition (i) of main text Sec.\ III. 

We first note that a pseudospike time $\tpseudo$ is smooth in case no network spike crosses the trial end. In this case, the network state at the trial end, i.e.\ potentials and currents at $T$, vary smoothly (\cref{sec:QIF ordinary proof}). Since $\tpseudo$ depends on the final network state via smooth functions (\cref{sec:QIF pseudo1}), it also varies smoothly. Thus, we only need to consider cases where a network spike crosses the trial end. As before, we only consider cases where only one spike crosses the trial end at a time.

\subsubsection{The spike crosses the trial end}\label{sec:QIF pseudo1 proof1}

We first show that the spike time changes smoothly with the network parameters, if a pseudospike becomes an ordinary spike or vice versa. Specifically, we consider the case where the $k$th spike crosses the trial end due to a small, continuous change of a network parameter. This means that at a critical value of this parameter, an ordinary spike (dis-)appears. If the parameter approaches the critical value from one side, the $k$th ordinary spike shifts towards the trial end, $\tspike\nearrow T$. This implies $V(T)\rightarrow-\infty$, since also the voltage reset following \tspike shifts towards the trial end from below. When approaching the critical parameter value from the other side, the spike and thus the voltage reset does not happen within the trial, but we have $V(T)\rightarrow\infty$, since the neuron comes closer to emitting its $k$th spike within the trial. In this case, the time of the $k$th spike, which is a pseudospike, is given by \cref{eq:QIF pseudo1 tpseudo} with $n_\trial=k-1$,
\begin{align}
    \tspike &= T + \phi_{\Theta,\Ipseudo} - \Phi_{\Ipseudo}(V(T)).
\end{align}
Because of $\lim\limits_{V(T)\rightarrow\infty} \Phi_{\Ipseudo} (V(T)) = \phi_{\Theta,\Ipseudo}$, $\lim\limits_{V(T)\rightarrow\infty} \tspike = T$. Thus, the pseudospike (dis-)appears at the trial end, where also the new ordinary spike (dis-)appears. This shows the continuity of the time of the $k$th spike in case it transitions from being an ordinary spike to being a pseudospike and vice versa.

To show that also the gradient is continuous, we consider a region in parameter space around the critical value for which, if the $k$th spike is a pseudospike, $V(T)$ is so large that the neuron would emit its $k$th spike if the trial would not end. We denote the time of this hypothetical ordinary spike by $\tord$, independent of the spike being before or after $T$. As established in \cref{sec:QIF ordinary proof}, \tord depends smoothly on the parameters. In particular, the value of its gradient at the transition is equal to its limit taken from either direction. It can be computed using \cref{eq:t analytic I=0,eq:QIF ext tspike,eq:QIF ext tspike w-} with $V_0=V(T)$ in case $\tord\gtrsim T$. The derivatives of \tord with respect to $V(T)$ and the input current at the trial end as well as the derivatives of \tpseudo with respect to $V(T)$ and \Ipseudo go to 0 when $V(T)\rightarrow\infty$. The derivative of $V(T)$ with respect to the varied parameter simultaneously diverges, however. Since the derivatives of \tord and \tpseudo with respect to $V(T)$ agree in leading order,
\begin{align}
    \diffp{\tpseudo}{V(T)} &= -\diffp{\Phi_{\Ipseudo(T)}(V(T))}{V(T)} = \frac{-1}{g(\Ipseudo) + (V(T)-1/2)^2} \underset{V(T)\rightarrow\infty}{\simeq} -\frac{1}{V(T)^2},\\
    \diffp{\tord}{V(T)} &\underset{V(T)\rightarrow\infty}{\simeq} -\frac{1}{V(T)^2},
\end{align}
also the gradients of \tord and \tpseudo asymptotically agree. Hence, the spike time gradient is continuous if the spike time crosses the trial end.

\subsubsection{A previous spike crosses the trial end}\label{sec:QIF pseudo1 proof2}

We now show that the spike time of a pseudospike changes smoothly with the network parameters, if a previous output spike of the same neuron crosses the trial end. Specifically, we consider the case where not the $k$th spike crosses the trial end but the $l$th spike, where $l<k$. When approaching the transition from the side where the $l$th spike is an ordinary spike, $n_\trial=l$ and $V(T)\rightarrow -\infty$. When approaching it from the other side, $n_\trial=l-1$ and $V(T)\rightarrow \infty$. In the former case, we have
\begin{align}
    \tpseudo &= T + (k-l)\phi_{\Theta,\Ipseudo} - \Phi_{\Ipseudo}(V(T))\underset{V(T)\rightarrow -\infty}{\longrightarrow} T + (k-l)\phi_{\Theta,\Ipseudo},
\end{align}
because $\lim\limits_{V(T)\rightarrow -\infty} \Phi_{\Ipseudo} (V(T)) = 0$. In the latter case, we have
\begin{align}
    \tpseudo &= T + (k-(l-1))\phi_{\Theta,\Ipseudo} - \Phi_{\Ipseudo}(V(T))\underset{V(T)\rightarrow \infty}{\longrightarrow} T + (k-l)\phi_{\Theta,\Ipseudo},
\end{align}
because $\lim\limits_{V(T)\rightarrow \infty} \Phi_{\Ipseudo} (V(T)) = \phi_{\Theta,\Ipseudo}$. Thus, $\tpseudo$ is continuous.

Since, in both cases,
\begin{align}
    \diffp{\tpseudo}{\Ipseudo} \rightarrow (k-l)\frac{-\pi}{2g^{(3/2)}(\Ipseudo)}\diffp{g(\Ipseudo)}{\Ipseudo}
\end{align}
and
\begin{align}
    \diffp{\tpseudo}{V(T)} = -\diffp{\Phi_{\Ipseudo}(V(T))}{V(T)} \simeq -\frac{1}{V(T)^2}
\end{align}
in leading order, also the gradient of $\tpseudo$ is continuous.

\subsubsection{An input spike crosses the trial end}\label{sec:QIF pseudo1 proof3}

Finally we show that the spike time of a pseudospike changes continuously with the network parameters, if a spike of another neuron in the network crosses the trial end. Specifically, we first assume that we move along a curve in parameter space that crosses a critical value where an input spike $k_{j_0}$ of neuron $j_0$ crosses the trial end. Since $V(T)$ changes continuously during the transition, we focus on $\Ipseudo$. When approaching the transition from the side where the $k_{j_0}$th spike of neuron $j_0$ is an ordinary spike, $I(T)\rightarrow I_{\mathstrike{k_{j_0}}}(T) + w_{j_0}$, where $I_{\mathstrike{k_{j_0}}}(T)$ is the value of the input current at the trial end without the effect of spike $k_{j_0}$. Furthermore \cref{eq:QIF pseudo1 r} implies $r_{j_0}\rightarrow 0$, because $V_{j_0}\searrow -\infty$. Thus, we have
\begin{align}
    \Ipseudo = I(T) + \sum_j w_j r_j \longrightarrow I_{\mathstrike{k_{j_0}}}(T) + w_{j_0} + \sum_{j\neq j_0} w_j r_j\,.
\end{align}
When approaching the transition from the other side, $I(T)\rightarrow I_{\mathstrike{k_{j_0}}}(T)$ and $r_{j_0}\rightarrow 1$. Thus, we have
\begin{align}
    \Ipseudo = I(T) + \sum_j w_j r_j \longrightarrow I_{\mathstrike{k_{j_0}}}(T) + \sum_{j\neq j_0} w_j r_j + w_{j_0}\,.
\end{align}
Since both limits agree, $\tpseudo$ is continuous at the transition. The continuity in case neuron $j_0$ is not directly presynaptic to neuron $i$ is then also guaranteed, since pseudospike times depend continuously on presynaptic pseudospike times.

The gradient of $\tpseudo$ is, however, not continuous in case an input spike crosses the trial end. 

\newpage

\section{Gradient statistics of QIFs with extended coupling}
\begin{figure}[h!]
    \centering
    \includegraphics{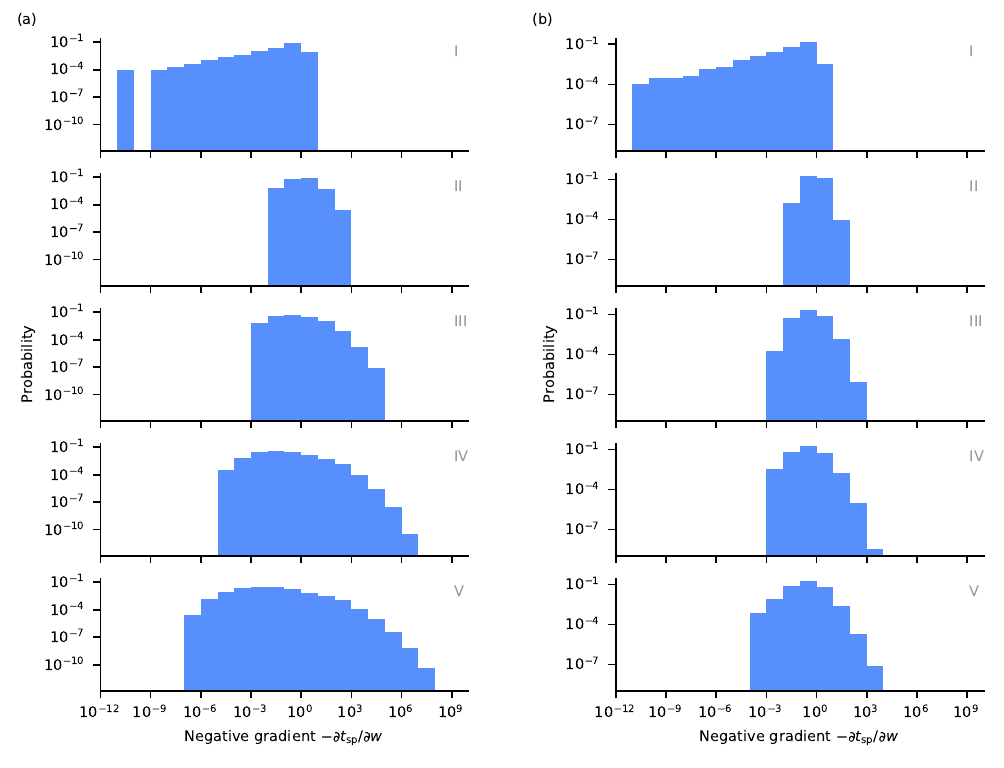}
    \caption{\label{fig:GradientStatistics}
    Spike time gradients of QIFs with extended coupling. (a) shows results for our standard and (b) for an intrinsically oscillating QIF, which moreover has lower input variance. We compute the gradients with respect to a test input weight and sample them according to the temporal distance of their underlying spike time to the test input (histograms I-V). 
    }\end{figure}

\begin{figure}[hb]
  \centering
  \includegraphics{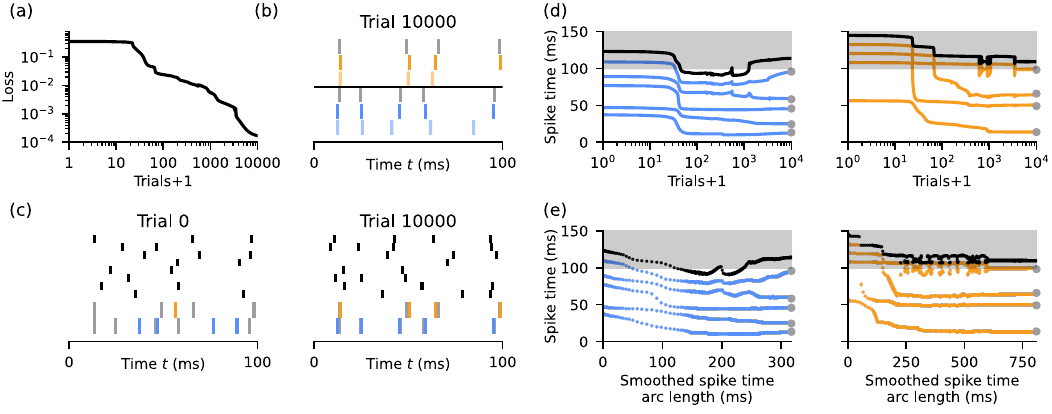}
  \caption{Further results on the learning of precise spikes in an RNN. (Same simulation as shown in main text Fig.~3.) 
  (a) Loss dynamics during learning. 
  (b) Comparison of target spike times (gray), spike times after learning (saturated colors) and spike times after learning if the weights of recurrent connections not targeting the first two neurons are set to 0 after learning (pale colors). The partially large deviations of the latter spike times from the targets illustrate that learning utilizes recurrent connections not involving the target neurons. 
  (c) (Reproduced from main text Fig.~3b.) Left: Spikes of network neurons before learning. Spikes of the first two neurons are colored, their target times are displayed in gray. Right: Learning changes the network dynamics such that the first two neurons spike precisely at the desired values (the colored spikes mostly cover the gray ones). 
  (d) Left: (Reproduced from main text Fig.~3c.) Evolution of the spike times of the first neuron during learning. The times of the spikes that are supposed to lie within the trial (blue traces) shift towards their target values (gray circles). The next spike (black trace) is supposed to lie outside the trial. It occurs transiently within the trial but becomes a pseudospike towards the end of learning again, as required. Gray area indicates pseudospikes. Right: Same as left but for the second neuron.
  (e) Same as (d) but the spike times are shown as a function of the arc length of the smoothed spike time trajectory. The spike times of the first neuron change continuously. The spike times of the second neuron exhibit jumps at which the times of later spikes shift to the times of earlier spikes at the previous trial. This is because of highly localized large gradients and can be avoided by using variable learning rates (see \cref{fig:fig3s2}). Furthermore, the spike times exhibit oscillations after the initial large shifts.
  }
  \label{fig:fig3s1}
\end{figure}

\begin{figure}
  \centering
  \includegraphics{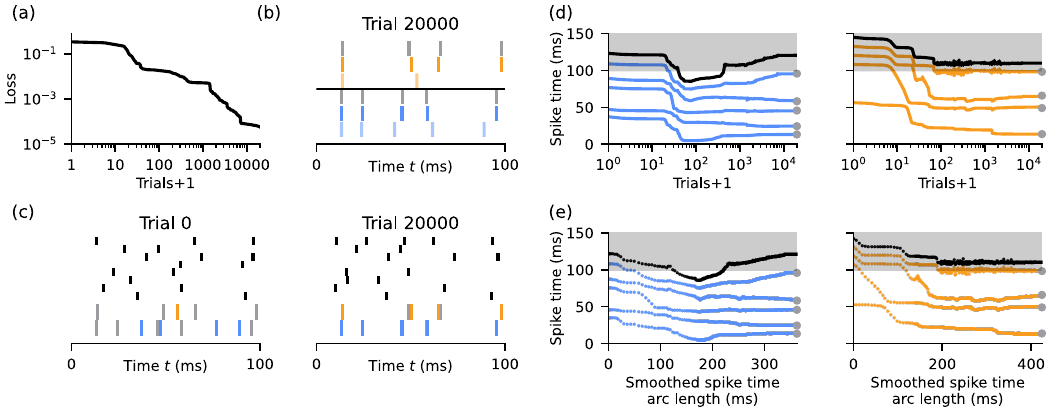}
  \caption{Same as \cref{fig:fig3s1} but using an alternative optimization method, which restricts the maximal step size (see \cref{tab:RNN description}). It results in continuous spike time changes (d,e), which reflects that the occurring gradients are large but finite.
  }
  \label{fig:fig3s2}
\end{figure}

In the following we numerically estimate magnitudes of the gradients that occur in QIFs with extended coupling. The neurons receive a high-frequency Poisson input spike train with normally distributed input weights. Inhibitory and excitatory spike inputs balance each other, such that the average input is zero. After a period of equilibration, a test input is provided. We compute the gradient with respect to the test input strength for different realizations of the input spike train and at different test input strengths. To cover the influence of temporal distance the obtained gradients are sorted according to the timing of the spike and presented in different histograms in \cref{fig:GradientStatistics}. Specifically, we bin time beyond the input into five bins of duration $2$ (two times the membrane time constant). The gradient of the time of a spike falling in bin number $n$ then contributes to the $n$th histogram (roman numerals in \cref{fig:GradientStatistics}). The $m$th bar in this histogram shows the empirical probability that in a single trial (with a randomly chosen test input weight and set of Poisson inputs) a spike time occurs in the $n$th time bin after the input and that it has a gradient that falls into the $m$th gradient size bin. The sum over these probabilities is the expected number of spikes per trial.

We observe that gradients of temporally close and of most distant spike times are often smaller than those of spikes with intermediate distance (compare histograms I,V with II,III,IV). This is because inputs usually have little impact on very close and very distant states. However, if a new spike (dis-)appears due to changes in the test input weight, this happens at the trial end, i.e.~with maximal temporal distance. These spikes have high sensitivity to the test weight as in the case without further inputs, cf.~the larger negative gradient around $w_{\min}$ in main text Fig.~1 left, which extends further for longer trial duration. Therefore the largest negative gradients occur in large temporal distance, \cref{fig:GradientStatistics}a IV and V. We find that both lower input variance and the addition of an oscillatory drive reduce the occurring gradients, \cref{fig:GradientStatistics}b. We finally note that for the standard exponential integrate-and-fire neuron with its steep upstroke towards spiking, we observe excessively large gradients already in very short trials.

\clearpage

\section{Further simulation results}

\begin{figure}
  \centering
  \includegraphics{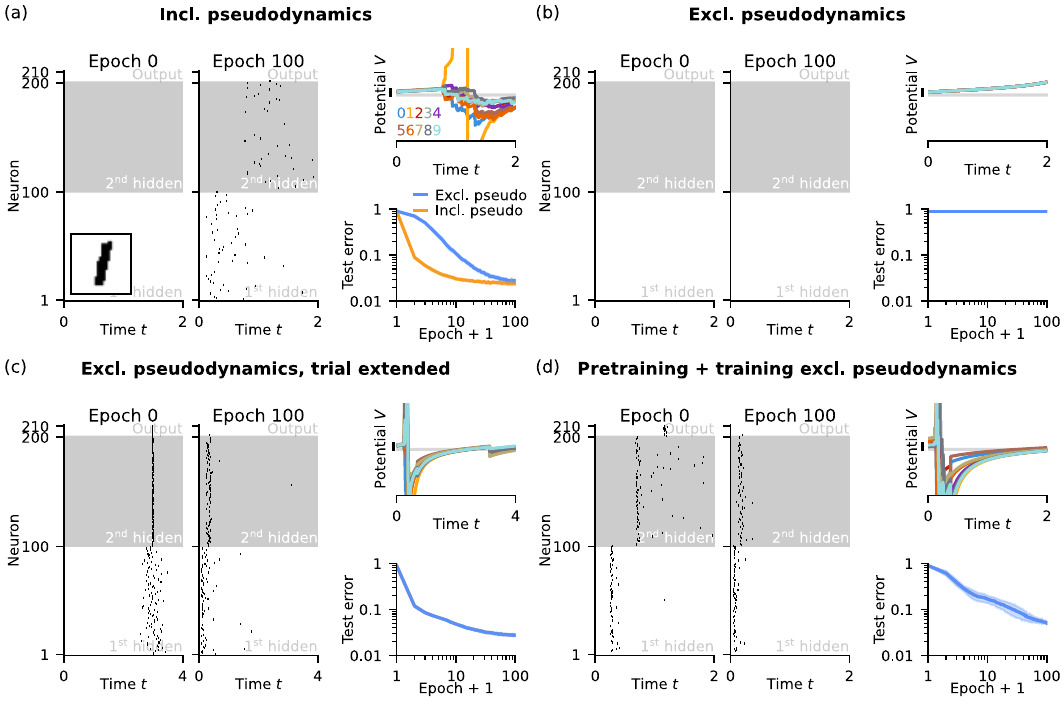}
  \caption{Necessity of pseudodynamics for the learning of the MNIST task. The panels show different learning variants. Each panel is structured like main text Fig.~4. The spike trains and voltage traces are evoked by the same stimulus, which is displayed as inset in (a). 
  (a) (Reproduced from main text Fig.~4.) Learning including pseudodynamics. (b) Learning excluding pseudodynamics. No new spikes can be added, hence learning is unsuccessful. (c) Learning excluding pseudodynamics but with an extended trial duration. Due to the extended trial and the oscillatory neuron model, all neurons already spike before learning, enabling successful learning. (d) Learning excluding pseudodynamics after pretraining. During pretraining, pseudodynamics are used to shift output neuron spikes inside the trial using an appropriate loss (see \cref{tab:mnist}). This loss is agnostic to the final task. After pretraining, the network neurons spike sufficiently often (leftmost panel) to enable successful learning. To compute the loss in panels (b-d), spike times of output neurons that do not spike within the trial are set to the trial duration $T$.
  }
  \label{fig:fig4s1}
\end{figure}

\clearpage

\section{Supplementary tables}

\begin{table}[ht]
\centering
\begin{minipage}{.47\textwidth}
    \centering
    \caption{Analysis of network behavior for the MNIST-task on the test set. Only spikes that lie within the trial and are relevant for the classification, i.e.~that happen before the first output spike, are considered. Specifically, for the computation of the accuracy, only ordinary output spikes are considered valid. For the loss, the spike times of output neurons that do not spike within the trial are set to $T$; this leads to a large loss also after training. A hidden neuron is considered silent if it does not spike before the first output spike (or, if there is no ordinary output spike, within the trial) for any input image. Similarly, the activity is the number of ordinary, hidden layer spikes before the first output spike (or, if there is no ordinary output spike, within the trial) per hidden neuron. Values represent mean $\pm$ std over ten network instances, clipped to lie within the possible ranges, where necessary. The standard deviation is zero for the accuracy and the loss before learning because there are no output spikes at all.}
    \label{tab:mnist results}
    \begin{tabular}{ccc}
        \toprule
         & {Before learning} & {After learning} \\
         \midrule
         Accuracy & \qty{9.8(0)}{\percent} & \qty{97.3(3)}{\percent} \\
         Loss & \num{2.320(0)} & \num{1.686(7)} \\
         Silent neurons & $(99.8\substack{+0.2 \\ -0.3})\si{\percent}$ & $(0.1\substack{+0.3 \\ -0.1})\si{\percent}$ \\
         Activity & $(1\substack{+2 \\ -1})\times 10^{-6}$ & \num{31.6(1)e-2}
    \end{tabular}
\end{minipage}
\hspace{.04\textwidth}
\begin{minipage}{.47\textwidth}
    \centering
    \caption{Same as \cref{tab:mnist results} but including the use of pseudospikes for classification and considering not only spikes before the first output spike but all ordinary spikes for the computation of the fraction of silent neurons and the activity.}
    \begin{tabular}{ccc}
        \toprule
         & {Before learning} & {After learning} \\
         \midrule
         Accuracy & \qty{10.4(7)}{\percent} & \qty{97.6(2)}{\percent} \\
         Loss & \num{2.345(4)} & \num{0.139(12)} \\
         Silent neurons & $(99.8\substack{+0.2 \\ -0.3})\si{\percent}$ & \qty{0(0)}{\percent} \\
         Activity & $(1\substack{+2 \\ -1})\times 10^{-6}$ & \num{41.6(12)e-2}
    \end{tabular}
\end{minipage}
\end{table}

\clearpage

\section{Supplementary discussion}

Our method and networks, in particular the one we use for the MNIST task, possess notable connections and parallels to standard and recent machine learning techniques. 
Probably the most direct connection is the fact that the working mode of our networks during the pseudodynamics of the first type corresponds to the functioning of a standard rate neural network. This means that during the pseudodynamics neurons interact via weighted sums of rate-like quantities, as shown in \cref{sec:QIF pseudo1,sec:QIF inf short pseudo}. A similar mapping can, however, not be obtained for the ordinary dynamics, even if there is at most one spike per neuron. This is because an ordinary spike time does not depend on the (possibly nonlinearly transformed) ordinary presynaptic spike times via their linear superposition, due to the nonlinear single neuron dynamics.
Thus, besides viewing our networks as fully spiking, one can also interpret them as a hybrid network, working in a spiking mode during the ordinary dynamics and in a rate mode during the pseudodynamics.
We note that there are mappings between networks of ReLUs and networks of carefully constructed non-leaky integrate-and-fire neurons that have step-like synaptic input currents and spike only once~\cite{stanojevic2023mapping,stanojevic2023highperformance}.

An important feature of our networks is that after learning, at inference time, it suffices to keep the ordinary dynamics only to achieve the tasks. In the setup that we use for the MNIST task, neurons typically only generate ordinary spikes in response to a subset of input images. Such sparsity is also important in machine learning contexts~\cite{hoefler2021sparsity}. It is a form of ephemeral (per example) sparsity, analogous to the sparsity induced by the ReLU activation function, which outputs zero for any negative (subthreshold) input, such that only an input-dependent subset of neurons is active for each input. 
During learning of the MNIST task, the interactions between pseudospikes imply that all neuron weights and activities affect the higher layer dynamics. This bears a similarity to using an activation function that does not clamp input ranges to zero, such as leaky ReLU or a related smooth function, in a non-spiking network during learning. The removal of the pseudodynamics after learning then resembles the replacement of this activation function by a ReLU function. This has been studied in~\cite{mirzadeh2024relu} in the context of large language models: 
Like in our networks the replacement sparsifies the network activity. It, however, also requires brief additional training. This is not the case in our networks, since pseudospikes do not affect ordinary ones and the ordinary spikes alone solve the task after training.

Pseudospikes allow the gradient to see behind the trial end. This bears some similarity to the property of a surrogate gradient to see under the threshold~\cite{Neftci2019surrogate}. In particular, both approaches can foresee whether spikes are about to appear for certain weight changes. However, in contrast to surrogate gradient descent, where the nonlinearities change, in our approach the computation of the dynamics (forward path) and the computation of the gradient (backward path) use the same dynamical equations. Further the contributions of the gradients of the ordinary spike times to the total gradient are not affected by the presence of pseudospikes.

A possible future research direction is the implementation of our approach on neuromorphic hardware. For this, it may be useful that with appropriately initialized network and trial parameters or after appropriate pretraining using pseudodynamics, training networks without pseudodynamics can be possible (\cref{fig:fig4s1}). If they are required, the pseudodynamics need special consideration depending on the learning scenario~\cite{christensen2022neuromorph}. This holds in particular for the pseudodynamics of the first type. They need at one point in time, namely at the ordinary trial end, the transmission of nonlinearly transformed subthreshold potentials between all connected neurons and the nonlinear computation of the strength of the constant input currents. In the case of on-chip learning, this may necessitate reading out the network state at $T$ and newly setting up the network with the appropriate currents and no interactions. The implementation of the second type of pseudodynamics may be easier: these dynamics first continue according to the same rules as the ordinary ones and at some fixed time point switch to different, fixed dynamics. In the case of off-chip learning, i.e.\ learning on conventional hardware and subsequent deployment of the learned network on neuromorphic hardware, the pseudodynamics do not cause a problem as they are not used after learning. If additional finetuning is required after deployment, adding spikes and thus pseudodynamics need not be necessary. In the case of chip-in-the-loop learning, i.e.\ neuromorphic hardware is used for the forward run and conventional hardware for the weight update computation, the pseudodynamics could be computed on the conventional hardware as well.

\clearpage

\section{Model and task details}
This section provides further details on the figures presented in our article. The formatting mostly follows ref.~\cite{nordlie2009modeldescription}. If not noted otherwise, the initial conditions are $V(0)=0$ (or the corresponding phase) and $I(0)=0$.

\begin{table}[hb]
\caption{Description of the QIF model of main text Fig.\ 1.}
\begin{tabular}{p{0.3\textwidth}p{0.6\textwidth}}
    \rowcolor{black}[0pt][5pt] {\textcolor{white}{\textbf{A}}}&{\textcolor{white}{\textbf{Model summary}}}\\
    \textbf{Population} & A single neuron\\
    \textbf{Neuron} & QIF\\
    \textbf{Synapse} & Extended coupling (exponentially decaying input current)\\
    \textbf{Input} & One or two input spikes
\end{tabular}

\begin{tabular}{p{0.3\textwidth}p{0.6\textwidth}}
    \rowcolor{black}[0pt][5pt] {\textcolor{white}{\textbf{B}}}&{\textcolor{white}{\textbf{Neuron and synapse model}}}\\
    \textbf{Name} & QIF with extended coupling\\
    \textbf{Neuron dynamics} & $\dot{V}(t)=V(t)(V(t)-1) + I(t)$ \hfill(Subthreshold dynamics)\\
    & $V_\Theta=\infty$ \hfill(Threshold)\\
    & $V_\reset=-\infty$ \hfill(Reset)\\
    \textbf{Synaptic dynamics} & $\taus \dot{I}(t)=-I(t) + \taus w\delta(t-t_\exc)$ \hfill(One input)\\
    & $\taus \dot{I}(t)=-I(t) + \taus w_\exc\delta(t-t_\exc) + \taus w_\inh\delta(t-t_\inh)$ \hfill(Two inputs)
\end{tabular}

\begin{tabular}{p{0.3\textwidth}p{0.6\textwidth}}
    \rowcolor{black}[0pt][5pt] {\textcolor{white}{\textbf{C}}}&{\textcolor{white}{\textbf{Input}}}\\
    \textbf{Type} & \textbf{Description}\\
    One input & A single excitatory input with varying weight $w$\\
    Two inputs & An excitatory input and an inhibitory input with varying time $t_\inh$
\end{tabular}

\begin{tabular}{p{0.15\textwidth}p{0.15\textwidth}p{0.6\textwidth}}
    \rowcolor{black}[0pt][5pt] {\textcolor{white}{\textbf{D}}}&&{\textcolor{white}{\textbf{Parameters}}}\\
    \textbf{Parameter} & \textbf{Value} & \textbf{Description}\\
    $T$ & $4$ & Trial length\\
    $\taus$ & $1/2$ & Synaptic time constant\\
    $w_\text{min}$ & $2.47$ & Minimal weight necessary to elicit a spike at infinity\\
    $t_\exc$ & $0.5$ & Time of excitatory input in both input cases\\
    $w_\exc$ & $1.5w_\text{min}$ & Weight of excitatory input in case of two inputs\\
    $w_\inh$ & $-w_\text{min}$ & Weight of inhibitory input in case of two inputs
\end{tabular}
\end{table}

\begin{table}
\caption{Description of the LIF model of main text Fig.\ 1.}
\begin{tabular}{p{0.3\textwidth}p{0.6\textwidth}}
    \rowcolor{black}[0pt][5pt] {\textcolor{white}{\textbf{A}}}&{\textcolor{white}{\textbf{Model summary}}}\\
    \textbf{Population} & A single neuron\\
    \textbf{Neuron} & LIF\\
    \textbf{Synapse} & Extended coupling (exponentially decaying input current)\\
    \textbf{Input} & One or two input spikes
\end{tabular}

\begin{tabular}{p{0.3\textwidth}p{0.6\textwidth}}
    \rowcolor{black}[0pt][5pt] {\textcolor{white}{\textbf{B}}}&{\textcolor{white}{\textbf{Neuron and synapse model}}}\\
    \textbf{Name} & LIF with extended coupling\\
    \textbf{Neuron dynamics} & $\dot{V}(t)= -V(t) + I(t)$ \hfill(Subthreshold dynamics)\\
    & $V_\Theta=1$ \hfill(Threshold)\\
    & $V_\reset=0$ \hfill(Reset)\\
    \textbf{Synaptic dynamics} & $\taus \dot{I}(t)=-I(t) + \taus w\delta(t-t_\exc)$ \hfill(One input)\\
    & $\taus \dot{I}(t)=-I(t) + \taus w_\exc\delta(t-t_\exc) + \taus w_\inh\delta(t-t_\inh)$ \hfill(Two inputs)
\end{tabular}

\begin{tabular}{p{0.3\textwidth}p{0.6\textwidth}}
    \rowcolor{black}[0pt][5pt] {\textcolor{white}{\textbf{C}}}&{\textcolor{white}{\textbf{Input}}}\\
    \textbf{Type} & \textbf{Description}\\
    One input & A single excitatory input with varying weight $w$\\
    Two inputs & An excitatory input and an inhibitory input with varying time $t_\inh$
\end{tabular}

\begin{tabular}{p{0.15\textwidth}p{0.15\textwidth}p{0.6\textwidth}}
    \rowcolor{black}[0pt][5pt] {\textcolor{white}{\textbf{D}}}&&{\textcolor{white}{\textbf{Parameters}}}\\
    \textbf{Parameter} & \textbf{Value} & \textbf{Description}\\
    $T$ & $3$ & Trial length\\
    $\taus$ & $1/2$ & Synaptic time constant\\
    $w_\text{min}$ & $4$ & Minimal weight necessary to elicit a spike\\
    $t_\exc$ & $0.5$ & Time of excitatory input in both input cases\\
    $w_\exc$ & $1.4w_\text{min}$ & Weight of excitatory input in case of two inputs\\
    $w_\inh$ & $-w_\text{min}$ & Weight of inhibitory input in case of two inputs
\end{tabular}
\end{table}

\begin{table}
\caption{Description of the QIF model of main text Fig.\ 2.}
\begin{tabular}{p{0.3\textwidth}p{0.6\textwidth}}
    \rowcolor{black}[0pt][5pt] {\textcolor{white}{\textbf{A}}}&{\textcolor{white}{\textbf{Model summary}}}\\
    \textbf{Population} & A single neuron\\
    \textbf{Neuron} & QIF\\
    \textbf{Synapse} & Extended coupling (exponentially decaying input current)\\
    \textbf{Input} & Combination of fixed, random as well as learnable input spikes\\
    \textbf{Learning} & Gradient descent on first two spike times
\end{tabular}

\begin{tabular}{p{0.3\textwidth}p{0.6\textwidth}}
    \rowcolor{black}[0pt][5pt] {\textcolor{white}{\textbf{B}}}&{\textcolor{white}{\textbf{Neuron and synapse model}}}\\
    \textbf{Name} & QIF with extended coupling\\
    \textbf{Neuron dynamics} & $\dot{V}(t)=V(t)(V(t)-1) + I(t)$ \hfill(Subthreshold dynamics)\\
    & $V_\Theta=\infty$ \hfill(Threshold)\\
    & $V_\reset=-\infty$ \hfill(Reset)\\
    \textbf{Synaptic dynamics} & $\taus \dot{I}(t)=-I(t)+\taus\sum\limits_{j=1}^{10}w_j^\fixed\delta(t-t_j^\fixed)+\taus\sum\limits_{j=1}^{2}w_j^\learn\delta(t-t_j^\learn)$\\
    \textbf{Pseudodynamics} & After the trial end, neurons evolve as described in \cref{sec:QIF pseudo1}
\end{tabular}

\begin{tabular}{p{0.3\textwidth}p{0.6\textwidth}}
    \rowcolor{black}[0pt][5pt] {\textcolor{white}{\textbf{C}}}&{\textcolor{white}{\textbf{Input}}}\\
    \textbf{Fixed inputs} & Twenty input spikes, times $t_j^\fixed$ randomly drawn from uniform distribution, weights $w_j^\fixed$ randomly drawn from normal distribution with mean $0$ and variance $1$ \\
    \textbf{Learnable inputs} & Two input spikes, times $t_j^\learn$ are initially $1$ and $9$, weights $w_j^\learn$ are initially $0$
\end{tabular}

\begin{tabular}{p{0.3\textwidth}p{0.6\textwidth}}
    \rowcolor{black}[0pt][5pt] {\textcolor{white}{\textbf{D}}}&{\textcolor{white}{\textbf{Learning}}}\\
    \textbf{Loss description} & Mean squared error loss\\
    \textbf{Loss function} & $L(p) = \frac{1}{2} \sum\limits_{k=1}^{2} \left(t_k(p)-t_k^\tar\right)^2$ \hfill($t_k(p)$ denotes the $k$th output spike)\\
    \textbf{Learnable parameters $p$} & Times $t_j^\learn$ and weights $w_j^\learn$ of the learnable input spikes\\
    \textbf{Optimization method} & Gradient descent with element-wise gradient clipping at \num{2.2e-2}
\end{tabular}

\begin{tabular}{p{0.15\textwidth}p{0.15\textwidth}p{0.6\textwidth}}
    \rowcolor{black}[0pt][5pt] {\textcolor{white}{\textbf{E}}}&&{\textcolor{white}{\textbf{Parameters}}}\\
    \textbf{Parameter} & \textbf{Value} & \textbf{Description}\\
    $T$ & $10$ & Trial length\\
    $\taus$ & $1/2$ & Synaptic time constant\\
    $\eta$ & $0.1$ &Learning rate\\
    $t_k^\tar$ & $\{2.5, 7.5\}$ & Target spike times\\
    $N_\trial$ & $3000$ & Number of trials
\end{tabular}
\end{table}

\begin{table}
\caption{Description of the RNN of main text Fig.\ 3.}\label{tab:RNN description}
\begin{tabular}{p{0.3\textwidth}p{0.6\textwidth}}
    \rowcolor{black}[0pt][5pt] {\textcolor{white}{\textbf{A}}}&{\textcolor{white}{\textbf{Model summary}}}\\
    \textbf{Population} & One population\\
    \textbf{Connectivity} & All-to-all\\
    \textbf{Neuron} & QIF\\
    \textbf{Synapse} & Extended coupling (exponentially decaying input current)\\
    \textbf{Input} & Excitatory and inhibitory Poisson spike trains\\
    \textbf{Learning} & Gradient descent on spike times of two network neurons
\end{tabular}

\begin{tabular}{p{0.3\textwidth}p{0.6\textwidth}}
     \rowcolor{black}[0pt][5pt] {\textcolor{white}{\textbf{B}}}&{\textcolor{white}{\textbf{Population}}}\\
     \multicolumn{2}{l}{One population of $N$ QIFs.}
\end{tabular}

\begin{tabular}{p{0.3\textwidth}p{0.6\textwidth}}
    \rowcolor{black}[0pt][5pt] {\textcolor{white}{\textbf{C}}}&{\textcolor{white}{\textbf{Connectivity}}}\\
    \addlinespace[0.2em]
    \multicolumn{2}{l}{\parbox{0.9\textwidth}{\raggedright{All-to-all recurrent connectivity, weights from neuron $j$ to neuron $i$ denoted with $w_{ij}$, weights initially set to 0}}}
\end{tabular}

\begin{tabular}{p{0.3\textwidth}p{0.6\textwidth}}
    \rowcolor{black}[0pt][5pt] {\textcolor{white}{\textbf{D}}}&{\textcolor{white}{\textbf{Neuron and synapse model}}}\\
    \textbf{Name} & QIF with extended coupling\\
    \textbf{Neuron dynamics} & $\dot{V}_i(t)=V_i(t)(V_i(t)-1) + I_i(t)$ \hfill(Subthreshold dynamics of neuron $i$)\\
    & $V_\Theta=\infty$ \hfill(Threshold)\\
    & $V_\reset=-\infty$ \hfill(Reset)\\
    \textbf{Synaptic dynamics} & $\taus \dot{I}_i(t)=-I_i(t)+\taus w_\exc^\text{in}S_{\exc,i}(t)+\taus w_\inh^\text{in}S_{\inh,i}(t) + \taus\sum\limits_{j=1, j\neq i}^N w_{ij} \sum\limits_{k_j}\delta(t-t_{k_j})$\\
    &\hfill($t_{k_j}$: $k$th spike time of neuron $j$)\\
    \textbf{Pseudodynamics} & After the trial end, neurons evolve as described in \cref{sec:QIF pseudo1}
\end{tabular}

\begin{tabular}{p{0.3\textwidth}p{0.6\textwidth}}
    \rowcolor{black}[0pt][5pt] {\textcolor{white}{\textbf{E}}}&{\textcolor{white}{\textbf{Input}}}\\
    \addlinespace[0.2em]
    \multicolumn{2}{l}{\parbox{0.9\textwidth}{\raggedright{Each neuron independently receives one excitatory Poisson spike train $S_{\exc,i}(t)=\sum_k \delta(t-t_k)$ with fixed weight $w_\exc^\text{in}$ and one inhibitory Poisson spike train $S_{\inh,i}(t)=\sum_k \delta(t-t_k)$ with fixed weight $w_\inh^\text{in}$. Both have the same rate $r_\text{in}$.}}}
\end{tabular}

\begin{tabular}{p{0.3\textwidth}p{0.6\textwidth}}
    \rowcolor{black}[0pt][5pt] {\textcolor{white}{\textbf{F}}}&{\textcolor{white}{\textbf{Learning}}}\\
    \textbf{Loss description} & Weighted mean squared error loss\\
    \textbf{Loss function} & \vspace{-2.5em}$$L(p) = \frac{1}{N_\target} \sum\limits_{i=1}^{N_\target}\sum\limits_{k_i=1}^{N_{\target,i}} \left(\frac{t_{k_i}(p)-t_{k_i}^\tar}{t_{k_i}^\tar+2}\right)^2 \left(1-\delta_{{k_i}N_{\target,i}}H(t_{N_{\target,i}}^\target-t_{k_i}(p))\right)$$\hfill($H$ is the Heaviside step function)\\
    \textbf{Learnable parameters $p$} & Initial states $V_i(0)$, $I_i(0)$ and recurrent weights $w_{ij}$ \\
    \textbf{Target times} & For $N_\target$ out of the $N$ network neurons, target times $t_{k_i}^\target$ are drawn from a Poisson process with rate $r_i^\target$ and absolute refractoriness 1. In addition to these $N_{\target,i}-1$ target spikes, a further target time $t_{N_{\target,i}}^\target=1.1T$ is used to avoid having more spikes than wanted within the trial.\\
    \textbf{Optimization method} & AdaBelief~\cite{zhuang2020adabelief} with exponential learning rate decay\\
    \textbf{Alternative optimization method} & AdaBelief~\cite{zhuang2020adabelief}, but with variable learning rate. In every step, the weight update is computed for a set of learning rates. Of all weight updates with a resulting maximal spike time change of less than $0.5$, the one resulting in the smallest error is selected.\\
\end{tabular}
\end{table}

\setcounter{table}{5}
\begin{table}
\caption{(continued)}
\begin{tabular}{p{0.15\textwidth}p{0.15\textwidth}p{0.6\textwidth}}
    \rowcolor{black}[0pt][5pt] {\textcolor{white}{\textbf{G}}}&&{\textcolor{white}{\textbf{Parameters}}}\\
    \textbf{Parameter} & \textbf{Value} & \textbf{Description}\\
    $T$ & $10$ & Trial length\\
    $N$ & $10$ & Number of neurons\\
    $\taus$ & $1/2$ & Synaptic time constant\\
    $w_\exc^\text{in}$ & $5$ & Excitatory input weight\\
    $w_\inh^\text{in}$ & $-w_\exc$ & Inhibitory input weight\\
    $r_\text{in}$ & $1$ & Input rate\\
    $r_1^\target$ & $1$ & Rate of the Poisson process used to generate target times for the first target neuron\\
    $r_2^\target$ & $1/2$ & Rate of the Poisson process used to generate target times for the second target neuron\\
    $N_\target$ & $2$ & Number of neurons whose spike times are learned\\
    $\eta$ & $0.01$ & Learning rate\\
    $\tau_\eta$ & \num{2e3} & Time scale of exponential learning rate decay (not used for the alternative optimization method)\\
    & $10^{-5}$--$10^2$ & Range of the 50 possible learning rates, evenly distributed in log-space, that are used in the alternative optimization method\\
    $\beta_1$ & $0.9$ & Exponential decay rate used to track first moment of gradient in AdaBelief\\
    $\beta_2$ & $0.999$ & Exponential decay rate used to track second moment of gradient in AdaBelief\\
    & $0.99$ & $\beta_2$ in the case of the alternative optimization method\\
    $N_\trial$ & \num{10000} & Number of trials\\
    & \num{20000} & $N_\trial$ in the case of the alternative optimization method
\end{tabular}
\end{table}

\begin{table}
\caption{\label{tab:mnist} Description of the multi-layer network of main text Fig.\ 4 and \cref{fig:fig4s1}.}
\begin{tabular}{p{0.3\textwidth}p{0.6\textwidth}}
    \rowcolor{black}[0pt][5pt] {\textcolor{white}{\textbf{A}}}&{\textcolor{white}{\textbf{Model summary}}}\\
    \textbf{Population} & Three: two hidden layers, one output layer\\
    \textbf{Connectivity} & Feedforward connectivity only\\
    \textbf{Neuron} & Oscillatory QIF\\
    \textbf{Synapse} & Infinitesimally short coupling (delta-pulse coupling)\\
    \textbf{Input} & Binarized MNIST images encoded with single spike per pixel\\
    \textbf{Learning} & Gradient descent learning of time-to-first spike encoded image label
\end{tabular}

\begin{tabular}{p{0.3\textwidth}p{0.6\textwidth}}
    \rowcolor{black}[0pt][5pt] {\textcolor{white}{\textbf{B}}}&{\textcolor{white}{\textbf{Population}}}\\
    \textbf{Input layer}& One input layer consisting of $N^{(0)}$ neurons with fixed spike times\\
    \textbf{Hidden layers}& Two hidden layers consisting of $N_\hid=N^{(1)}=N^{(2)}$ neurons each\\
    \textbf{Output layer}& One output layer consisting of $N^{(3)}=N_\target=10$ neurons, one for each label
\end{tabular}

\begin{tabular}{p{0.3\textwidth}p{0.6\textwidth}}
    \rowcolor{black}[0pt][5pt] {\textcolor{white}{\textbf{C}}}&{\textcolor{white}{\textbf{Connectivity}}}\\
    \addlinespace[0.2em]
    \multicolumn{2}{l}{\parbox{0.9\textwidth}{\raggedright{Full feedforward connectivity between subsequent layers, no recurrent connections, weight from neuron $j$ in layer $l-1$ to neuron $i$ in layer $l$ denoted with $w^{(l)}_{ij}$, weights initially randomly drawn from uniform distribution}}}
\end{tabular}

\begin{tabular}{p{0.3\textwidth}p{0.6\textwidth}}
    \rowcolor{black}[0pt][5pt] {\textcolor{white}{\textbf{D}}}&{\textcolor{white}{\textbf{Neuron and synapse model}}}\\
    \textbf{Name} & Oscillatory QIF with delta-pulse coupling\\
    \textbf{Neuron dynamics} & $\taum\dot{V}_i^{(l)}(t)=V_i^{(l)}(t)(V_i^{(l)}(t)-1) + I_i^{(l)}(t)$\\
    &\hfill(Subthreshold dynamics  of neuron $i$ in layer $l$)\\
    & $V_\Theta=\infty$ \hfill(Threshold)\\
    & $V_\reset=-\infty$ \hfill(Reset)\\
    \textbf{Synaptic dynamics} & $I_i^{(l)}(t)=I_0 + \taum\sum\limits_{j=1}^{N^{(l-1)}} w_{ij}^{(l)} \sum\limits_{k_j}\delta(t-t_{k_j})$\hfill($t_{k_j}$: $k$th spike time of neuron $j$)\\
    \textbf{Neuron dynamics (angle space)} & $\dot{\phi}_i^{(l)}(t)=1$ \hfill(Between spikes)\\
    & $\phi_\Theta=\taum\pi/\sqrt{I_0-\frac{1}{4}}$ \hfill(Threshold)\\
    & $\phi_\reset=0$ \hfill(Reset)\\
    \textbf{Synaptic dynamics (angle space)} & $\phi_i^{(l)}(t_{k_j}^+)=H_{w_{ij}^{(l)}}(\phi_i^{(l)}(t_{k_j}^-)) =\Phi(\Phi^{-1}(\phi_i^{(l)}(t_{k_j}^-))+w_{ij}^{(l)})$\\
    & \hfill($t_{k_j}$ denotes the $k$th spike of neuron $j$)\\
    \textbf{Pseudodynamics} & After the trial end, neurons evolve as described in \cref{sec:QIF inf short pseudo}
\end{tabular}

\begin{tabular}{p{0.3\textwidth}p{0.6\textwidth}}
    \rowcolor{black}[0pt][5pt] {\textcolor{white}{\textbf{E}}}&{\textcolor{white}{\textbf{Input}}}\\
    \addlinespace[0.2em]
    \multicolumn{2}{l}{\parbox{0.9\textwidth}{\raggedright{Pixel values are binarized, input neurons corresponding to active pixels spike once at $0.02$, others do not spike at all}}}
\end{tabular}
\end{table}

\setcounter{table}{6}
\begin{table}
\caption{(continued)}
\begin{tabular}{p{0.3\textwidth}p{0.6\textwidth}}
    \rowcolor{black}[0pt][5pt] {\textcolor{white}{\textbf{F}}}&{\textcolor{white}{\textbf{Learning}}}\\
    \textbf{Loss description} & Cross-entropy loss on first spike times of the output neurons, regularization term to encourage early spiking~\cite{Göltz2021neuromorphic}\\
    \textbf{Loss function (single input)} & \vspace{-1em}\parbox{0.6\textwidth}{
    $$\begin{aligned}
        L(p) &=  \sum\limits_{i=1}^{N_\target}y_{\target,i}\log(y_i(p)) + \gamma \sum\limits_{i=1}^{N_\target}y_{\target,i} \left(\exp(t_i^{(3)}(p)/T)-1\right)\\
        y_i(p) &= \frac{\exp(-t_i^{(3)}(p))}{\sum_j \exp(-t_j^{(3)}(p))}\hspace{.3\textwidth}\text{(softmax)}\\
        y_{\target,i} &= \delta_{i,\text{(label+1)}}\hspace{.2\textwidth}\text{(one-hot encoded target label)}
    \end{aligned}$$
    }\\
    \textbf{Loss function (pretraining in \cref{fig:fig4s1}(d))} & \vspace{-1em}\parbox{0.6\textwidth}{
    $$\begin{aligned}
        L(p) &=  \frac{1}{N_\target}\sum\limits_{i=1}^{N_\target}\left(t_i^{(3)}-T\right)^2 H\left(t_i^{(3)}-T\right)
    \end{aligned}$$
    \hfill($H$ is the Heaviside step function)
    }\\
    \textbf{Learnable parameters $p$} & Initial states $V_i^{(l)}(0)$ and feedforward weights $w_{ij}^{(l)}$ \\
    \textbf{Mini batches} &  Batches of size $N_\batch$ are used, loss is averaged over batch\\
    \textbf{Optimization method} & AdaBelief~\cite{zhuang2020adabelief} with exponential learning rate decay\\
    \textbf{Input regularization} & To avoid overfitting, the state of each binarized pixel is flipped with probability $p_\flip$ during learning\\
    \textbf{Hyperparameter search and evaluation} & Training data set: $55000$ images, validation data set: $5000$ images, test data set: $10000$, hyperparameters are manually tuned using the validation data set, network performance is evaluated on held-out test data set
\end{tabular}

\begin{tabular}{p{0.1\textwidth}p{0.2\textwidth}p{0.6\textwidth}}
    \rowcolor{black}[0pt][5pt] {\textcolor{white}{\textbf{G}}}&&{\textcolor{white}{\textbf{Parameters}}}\\
    \textbf{Parameter} & \textbf{Value} & \textbf{Description}\\
    $T$ & $2$ & Trial length\\
    $N^{(0)}$ & $784$ & Number of input neurons/pixels\\
    $N_\hid$ & $100$ & Number of hidden layer neurons\\
    $N_\target$ & $10$ & Number of output neurons\\
    $\taum$ & $6/\pi$ & Membrane time constant\\
    $I_0$ & $5/4$ & Constant input current component\\
    & $\mathcal{U}([\frac{-0.5}{\sqrt{N^{(l-1)}}},\frac{0.5}{\sqrt{N^{(l-1)}}}])$ & Distribution of weights $w_{ij}^{(l)}$ before learning\\
    & $\phi_\Theta / 2$ & Value of initial states $V_i^{(l)}(0)$ before learning\\
    $\gamma$ & $10^{-2}$ & Regularization parameter\\
    $N_\batch$ & $1000$ & Batch size\\
    $\eta$ & $4\times 10^{-3}$ & Learning rate\\
    $\tau_\eta$ & $10^{2}$ & Time scale of exponential learning rate decay\\
    $\beta_1$ & $0.9$ & Exponential decay rate used to track first moment of gradient in  AdaBelief\\
    $\beta_2$ & $0.999$ & Exponential decay rate used to track second moment of gradient in  AdaBelief\\
    $p_\flip$ & $0.02$ & Flip probability of each pixel during learning\\
    $N_\epoch$ & $100$ & Number of epochs (passes through the entire training data set) used for learning\\
    & $1$ & Number of epochs used for pretraining in \cref{fig:fig4s1}(d).
\end{tabular}
\end{table}

\clearpage

\begin{table}[hb]
\caption{Description of the multi-layer network of \cref{fig:Pseudospikes type 2}.
}
\begin{tabular}{p{0.3\textwidth}p{0.6\textwidth}}
    \rowcolor{black}[0pt][5pt] {\textcolor{white}{\textbf{A}}}&{\textcolor{white}{\textbf{Model summary}}}\\
    \textbf{Population} & Two: One hidden layer, one output layer\\
    \textbf{Connectivity} & Feedforward connectivity only\\
    \textbf{Neuron} & QIF\\
    \textbf{Synapse} & Extended coupling (exponentially decaying input current)\\
    \textbf{Input} & Two input spikes
\end{tabular}

\begin{tabular}{p{0.3\textwidth}p{0.6\textwidth}}
    \rowcolor{black}[0pt][5pt] {\textcolor{white}{\textbf{B}}}&{\textcolor{white}{\textbf{Population}}}\\
    \textbf{Input layer}& One input layer consisting of two neurons with fixed spike times\\
    \textbf{Hidden layers}& One hidden layer consisting of two neurons\\
    \textbf{Output layer}& One output layer consisting of one neuron
\end{tabular}

\begin{tabular}{p{0.3\textwidth}p{0.6\textwidth}}
    \rowcolor{black}[0pt][5pt] {\textcolor{white}{\textbf{C}}}&{\textcolor{white}{\textbf{Connectivity}}}\\
    \addlinespace[0.2em]
    \multicolumn{2}{l}{\parbox{0.9\textwidth}{\raggedright{Input neuron 1 excites both hidden neurons with the same variable synaptic strength $w$, $w^1_{11}=w^1_{21}=w$.
    Input neuron 2 has no connection to hidden neuron 1, $w^1_{12}=0$, and inhibits
hidden neuron 2 with fixed synaptic strength $w^1_{22}=-2$.  Hidden neuron 1 excites the output neuron with fixed synaptic strength $w^2_{11}=3$. Hidden neuron 2 inhibits the output neuron with fixed synaptic strength $w^2_{12}=-1$. $w$ changes from $2$ to $6$ in steps of $\Delta w=10^{-6}$. We compute the spike time gradient with respect to the synaptic weight $w$ using the change of same spike times $\Delta \tspike$ between subsequent $w$ as  $\Delta \tspike/\Delta w$.}}}
\end{tabular}

\begin{tabular}{p{0.3\textwidth}p{0.6\textwidth}}
    \rowcolor{black}[0pt][5pt] {\textcolor{white}{\textbf{D}}}&{\textcolor{white}{\textbf{Neuron and synapse model}}}\\
    \textbf{Name} & QIF with extended coupling\\
    \textbf{Neuron dynamics (angle space)} & 
    $\dot{\phi}^{(l)}_i(t)=\cos(\pi\phi^{(l)}_i(t))\left(\cos(\pi\phi^{(l)}_i(t))+\frac{1}{\pi}\sin(\pi\phi^{(l)}_i(t))\right)$\\&$\phantom{\dot{\phi}^{(l)}_i(t)}+\frac{1}{\pi^2}\sin^2(\pi\phi^{(l)}_i(t))I^{(l)}_i(t)$
    \\&\hfill(Subthreshold dynamics of neuron $i$ in layer $l$)\\
    & $\phi_\Theta=1$ \hfill(Threshold)\\
    & $\phi_\reset=0$ \hfill(Reset)\\
    \textbf{Synaptic dynamics} & $\taus \dot{I}^{(l)}_i(t)=-I^{(l)}_i(t)+\taus\sum\limits_{j=1}^{2} w^{(l)}_{ij} \sum\limits_{k_j}\delta(t-t_{k_j})$\\&\hfill($t_{k_j}$: $k$th spike time of neuron $j$)\\
    \textbf{Pseudodynamics} & After the trial end, neurons evolve as described in \cref{sec:QIF pseudo2}
\end{tabular}

\begin{tabular}{p{0.3\textwidth}p{0.6\textwidth}}
    \rowcolor{black}[0pt][5pt] {\textcolor{white}{\textbf{E}}}&{\textcolor{white}{\textbf{Input}}}\\
    \textbf{Type} & \textbf{Description}\\
    Input spikes & One input spike from input layer neuron 1 at time $t=1$. One input spike from input layer neuron 2 at time $t=0$.
\end{tabular}

\begin{tabular}{p{0.15\textwidth}p{0.15\textwidth}p{0.6\textwidth}}
    \rowcolor{black}[0pt][5pt] {\textcolor{white}{\textbf{F}}}&&{\textcolor{white}{\textbf{Parameters}}}\\
    \textbf{Parameter} & \textbf{Value} & \textbf{Description}\\
    $T$ & $8$ & Trial length \\
    $d$ & $2$ & Pseudospike dynamics parameter, \cref{eq:Tl pseudo 2}
\end{tabular}
\end{table}

\begin{table}[hb]
\caption{Description of the QIF model of \cref{fig:tsp(w)}. 
}
\begin{tabular}{p{0.3\textwidth}p{0.6\textwidth}}
    \rowcolor{black}[0pt][5pt] {\textcolor{white}{\textbf{A}}}&{\textcolor{white}{\textbf{Model summary}}}\\
    \textbf{Population} & A single neuron\\
    \textbf{Neuron} & QIF\\
    \textbf{Synapse} & Extended coupling (exponentially decaying input current)\\
    \textbf{Input} & One test input
\end{tabular}

\begin{tabular}{p{0.3\textwidth}p{0.6\textwidth}}
    \rowcolor{black}[0pt][5pt] {\textcolor{white}{\textbf{B}}}&{\textcolor{white}{\textbf{Neuron and synapse model}}}\\
    \textbf{Name} & QIF with extended coupling\\
    \textbf{Neuron dynamics (angle space)} & 
    $\dot{\phi}(t)=\cos(\pi\phi(t))\left(\cos(\pi\phi(t))+\frac{1}{\pi}\sin(\pi\phi(t))\right)$\\&$\phantom{\dot{\phi}(t)}+\frac{1}{\pi^2}\sin^2(\pi\phi(t))I(t)$
    \\&\hfill(Subthreshold dynamics of neuron $i$ in layer $l$)\\
    & $\phi_\Theta=1$ \hfill(Threshold)\\
    & $\phi_\reset=0$ \hfill(Reset)\\
    \textbf{Synaptic dynamics} & $\taus \dot{I}(t)=-I(t)$\\&\hfill($I(0)=w$: weight of test input)
\end{tabular}

\begin{tabular}{p{0.3\textwidth}p{0.6\textwidth}}
    \rowcolor{black}[0pt][5pt] {\textcolor{white}{\textbf{C}}}&{\textcolor{white}{\textbf{Input}}}\\
    \textbf{Type} & \textbf{Description}\\
    Test input & Input time $t=0$, input weight $w$ varied between $w_{\min}=-8.5$ and $w_{\max}=60$ in steps of $10^{-4}$.
\end{tabular}

\begin{tabular}{p{0.15\textwidth}p{0.15\textwidth}p{0.6\textwidth}}
    \rowcolor{black}[0pt][5pt] {\textcolor{white}{\textbf{D}}}&&{\textcolor{white}{\textbf{Parameters}}}\\
    \textbf{Parameter} & \textbf{Value} & \textbf{Description}\\
    $T$ & $10$ & Trial length\\
    $\phi(0)$
    &  $\Phi(3)\approx 0.74$ &Initial phase
\end{tabular}
\end{table}

\begin{table}[hb]
\caption{Description of the QIF model of \cref{fig:tsp(ti),fig:tsp(wi)}.}
\begin{tabular}{p{0.3\textwidth}p{0.6\textwidth}}
    \rowcolor{black}[0pt][5pt] {\textcolor{white}{\textbf{A}}}&{\textcolor{white}{\textbf{Model summary}}}\\
    \textbf{Population} & A single neuron\\
    \textbf{Neuron} & QIF\\
    \textbf{Synapse} & Extended coupling (exponentially decaying input current)\\
    \textbf{Input} & One test input, four further inputs.
\end{tabular}

\begin{tabular}{p{0.3\textwidth}p{0.6\textwidth}}
    \rowcolor{black}[0pt][5pt] {\textcolor{white}{\textbf{B}}}&{\textcolor{white}{\textbf{Neuron and synapse model}}}\\
    \textbf{Name} & QIF with extended coupling\\
    \textbf{Neuron dynamics (angle space)} & 
    $\dot{\phi}(t)=\cos(\pi\phi(t))\left(\cos(\pi\phi(t))+\frac{1}{\pi}\sin(\pi\phi(t))\right)$\\&$\phantom{\dot{\phi}(t)}+\frac{1}{\pi^2}\sin^2(\pi\phi(t))I(t)$
    \\&\hfill(Subthreshold dynamics of neuron $i$ in layer $l$)\\
    & $\phi_\Theta=1$ \hfill(Threshold)\\
    & $\phi_\reset=0$ \hfill(Reset)\\
    \textbf{Synaptic dynamics} & $\taus \dot{I}(t)=-I(t)+\taus\sum\limits_{j} w_{j} \delta(t-t_j)$\\&\hfill($w_j,t_j$: weight and time of $j$th input)
\end{tabular}

\begin{tabular}{p{0.3\textwidth}p{0.6\textwidth}}
    \rowcolor{black}[0pt][5pt] {\textcolor{white}{\textbf{C}}}&{\textcolor{white}{\textbf{Input}}}\\
    \textbf{Type} & \textbf{Description}\\
    Test input & \cref{fig:tsp(ti)}: Input time $t_i$ varied between $0$ and $8$ in steps of $10^{-5}$, input weight $w_i=-3$.
    \\&\cref{fig:tsp(wi)}: Input time $t_i=2.22$, input weight $w_i$ varied between $w_{\min}=-8.5$ and $w_{\max}=60$ in steps of $10^{-4}$.\\
    Further inputs & Input from input neuron $1$ at times $1$ and $1.5$, weight $2$. Input from input neuron $2$ at time $3$, weight $-3$. Input from input neuron $3$ at time $5$, weight $4$.
\end{tabular}

\begin{tabular}{p{0.15\textwidth}p{0.15\textwidth}p{0.6\textwidth}}
    \rowcolor{black}[0pt][5pt] {\textcolor{white}{\textbf{D}}}&&{\textcolor{white}{\textbf{Parameters}}}\\
    \textbf{Parameter} & \textbf{Value} & \textbf{Description}\\
    $T$ & $10$ & Trial length
\end{tabular}

\end{table}

\begin{table}[hb]
\caption{Description of the QIF models and the analysis of \cref{fig:GradientStatistics}.}
\begin{tabular}{p{0.3\textwidth}p{0.6\textwidth}}
    \rowcolor{black}[0pt][5pt] {\textcolor{white}{\textbf{A}}}&{\textcolor{white}{\textbf{Model summary}}}\\
    \textbf{Population} & A single neuron\\
    \textbf{Neuron} & (a) QIF, (b) Oscillatory QIF\\
    \textbf{Synapse} & Extended coupling (exponentially decaying input current)\\
    \textbf{Input} & One test input, balanced Poisson spike train
\end{tabular}

\begin{tabular}{p{0.3\textwidth}p{0.6\textwidth}}
    \rowcolor{black}[0pt][5pt] {\textcolor{white}{\textbf{B}}}&{\textcolor{white}{\textbf{Neuron and synapse model}}}\\
    \textbf{Name} & QIF with extended coupling\\
    \textbf{Neuron dynamics} & $\dot{V}(t)=V(t)(V(t)-1) + I(t)$ \hfill(Subthreshold dynamics)\\
    & $V_\Theta=10000$ \hfill(Threshold)\\
    & $V_\reset=-10000$ \hfill(Reset)\\
    \textbf{Synaptic dynamics} & $\taus \dot{I}(t)=-I(t)+I_0+\taus\sum\limits_{i} w_{i} \delta(t-t_i)$\\
    &\hfill($w_i,t_i$: weight and time of $i$th input spike)
\end{tabular}

\begin{tabular}{p{0.3\textwidth}p{0.6\textwidth}}
    \rowcolor{black}[0pt][5pt] {\textcolor{white}{\textbf{C}}}&{\textcolor{white}{\textbf{Input}}}\\
    \textbf{Type} & \textbf{Description}\\
        Test input & Input time $t=5$, uniformly distributed, strength between $w_{\min}=-1.5$ and $w_{\max}=1.5$.\\
    Poisson input & Input arrivals during the entire trial, frequency $10$ ($1$kHz), weights $w_i$ normally distributed, standard deviation (a) $\sigma=0.5$, (b) $\sigma=0.25$.
\end{tabular}

\begin{tabular}{p{0.15\textwidth}p{0.15\textwidth}p{0.6\textwidth}}
    \rowcolor{black}[0pt][5pt] {\textcolor{white}{\textbf{D}}}&&{\textcolor{white}{\textbf{Parameters}}}\\
    \textbf{Parameter} & \textbf{Value} & \textbf{Description}\\
    $T$ & $15$ & Trial length\\
    $I_0$
    & (a) $0$, (b) $0.5$ &Constant drive
\end{tabular}

\begin{tabular}{p{0.3\textwidth}p{0.6\textwidth}}
    \rowcolor{black}[0pt][5pt] {\textcolor{white}{\textbf{E}}}&{\textcolor{white}{\textbf{Analysis}}}\\
    \textbf{Type} & \textbf{Description}\\
    Empirical probability estimation & We consider 10000 sets of trials. In a single set, the randomly chosen Poisson input spike trains are kept the same, while we sample the test input. To resolve also steep gradients, we use an adaptive sampling scheme: The test input weight is decreased from $w_{\max}$ to $w_{\min}$ with an initial and (in absolute value) maximal step size of $\Delta w=-0.1$. The spike times thus increase between subsequent trials. We choose as desired maximum of the spike time differences $\Delta \tspike$ between same spikes in trial $i+1$ and $i$ the value $0.1$. If it is exceeded by a factor of $2$, trial $i+1$ is discarded and $\Delta w$ is reduced by a factor of $2$. If the observed maximum is smaller than desired by a factor of $2$, $\Delta w$ is increased by a factor of $2$, up to the maximal step size. We compute the negative gradients via $-\Delta \tspike/\Delta w$. After the trial set is completed, we sum the lengths of the test weight intervals for which a spike lies in time bin $n$ (bin size $2$) and has a gradient in size bin $m$. The result is normalized by the entire test weight interval sampled. This gives the trial set's probability estimate for bin $m$ in histogram $n$. Averaging over all trial sets yields the final result. 
\end{tabular}

\end{table}
\clearpage

\providecommand{\noopsort}[1]{}\providecommand{\singleletter}[1]{#1}%